\documentclass[aps,prd,twocolumn,preprintnumbers,amsmath,amssymb,floatfix,nofootinbib]{revtex4-1}

\usepackage{dcolumn}
\usepackage{bm}
\usepackage[dvips]{graphicx}
\usepackage{amsmath}
\usepackage{epsfig}
\usepackage{amsfonts}
\usepackage{amssymb}

\newcommand{\be}{\begin{equation}}
\newcommand{\ee}{\end{equation}}
\newcommand{\bee}{\begin{equation*}}
\newcommand{\eee}{\end{equation*}}
\newcommand{\bea}{\begin{eqnarray}}
\newcommand{\eea}{\end{eqnarray}}
\newcommand{\bean}{\begin{eqnarray*}}
\newcommand{\eean}{\end{eqnarray*}}

\usepackage{color}

\newcommand{\cba}{\ensuremath{c_{\beta-\alpha}}}
\newcommand{\sba}{\ensuremath{s_{\beta-\alpha}}}

\begin{document}

\preprint{DESY 15-240}

\title{Hierarchical vs Degenerate 2HDM: \\The LHC Run 1 Legacy at the Onset of Run 2}

\author{G. C. Dorsch$^{1,2}$, S. J. Huber$^{1}$, K. Mimasu$^{1}$ and J. M. No$^{1}$}

\affiliation{$^{1}${\it Department of Physics and Astronomy, University of Sussex, 
BN1 9QH Brighton, United Kingdom}}
\affiliation{$^{2}${\it DESY, Notkestrasse 85, D-22607 Hamburg, Germany}}

\date{\today}

\begin{abstract}
Current discussions of the allowed two-Higgs-doublet model (2HDM) parameter space after LHC Run 1 and the prospects for Run 2 are commonly phrased in the context of a 
quasi-degenerate spectrum for the new scalars. Here we discuss the generic situation of a 2HDM with a non-degenerate spectrum for the new scalars. 
This is highly motivated from a cosmological perspective since it naturally leads to a strongly first order electroweak phase transition that could 
explain the matter-antimatter asymmetry in the Universe. While constraints from measurements of Higgs signal strengths do not change, those from searches of 
new scalar states get modified dramatically once a non-degenerate spectrum is considered. 
\end{abstract}

\maketitle

\section{Introduction}
\label{sec:intro}

\vspace{-3mm}

While ongoing analyses by both ATLAS and CMS show that the properties of the Higgs particle at $m_h\sim$~125~GeV
are close to those expected for the Standard Model (SM) Higgs boson $h_{\mathrm{SM}}$~\cite{Aad:2015gba,Khachatryan:2014jba}, 
the complete nature of the scalar sector responsible for electroweak (EW) symmetry-breaking 
remains to be determined. It is particularly interesting to ascertain whether the scalar sector consists of only one $\mathrm{SU(2)}_L$ doublet or
has a richer structure containing additional states. Addressing this question is a very important task for present and future studies at the Large Hadron Collider (LHC).

In this work we concentrate on models with two Higgs doublets (2HDM) (see~\cite{Branco:2011iw} for a review), which appear in many extensions of the SM, such as the MSSM or
scenarios of viable Electroweak Baryogenesis~\cite{Cline:1995dg,Cline:1996mga,Fromme:2006cm,Dorsch:2013wja,Dorsch:2014qja}. 
In recent years, the region of the parameter space for the 2HDM allowed by Higgs coupling measurements at ATLAS and CMS~\cite{Aad:2015pla} has been widely 
studied in the literature~\cite{Celis:2013rcs,Krawczyk:2013gia,Grinstein:2013npa,Chen:2013rba,Craig:2013hca,Eberhardt:2013uba,Dumont:2014wha,Bernon:2014vta}.
Various works have also discussed the constraints coming from LHC searches for neutral and charged scalars $A_0$, $H_0$, $H^{\pm}$ via
$A_0 \to Z h$, $A_0/H_0 \to \tau\tau$, $A_0/H_0 \to \gamma\gamma$, $H_0 \to Z Z /W W$, $H_0 \to h h$, $H^{\pm} \to tb$ and $H^{\pm} \to \tau \nu$ 
(see {\it e.g.}~\cite{Craig:2015jba,Bernon:2015qea} for recent analyses). However, the interpretation of these constraints typically assumes 
that the new scalars can only decay into SM states, which requires $H_0$, $A_0$ and $H^{\pm}$ to be relatively close in 
mass (see however~\cite{Coleppa:2014hxa,Coleppa:2014cca,Li:2015lra}). 
In the following we refer to this scenario as the {\it degenerate} 2HDM. 

On the other hand, it has recently been shown~\cite{Dorsch:2014qja} that sizable mass splittings between the 
2HDM scalars (in particular a large $m_{A_0} - m_{H_0}$) favour a strong EW phase transition that could lead to baryogenesis. This provides an important physical motivation 
for a 2HDM scenario  in which new decay channels for the heavier scalars are kinematically allowed ({\it e.g.} $A_0 \to Z H_0$), 
a situation which has so far been largely neglected in the literature. We refer to this scenario 
as the {\it hierarchical} 2HDM. It is the purpose of this work to fill this gap, providing a detailed discussion of the constraints on the 2HDM 
parameter space from 7 and 8 TeV LHC Run 1 data, comparing the degenerate and hierarchical 2HDM scenarios. 
We will show that, besides significantly weakening the bounds from searches for these new scalars into 
SM states, the sizable mass splittings provide possibilities for novel searches (see {\it e.g.}~\cite{CMS:2015mba}) which can yield 
complementary limits on the 2HDM parameter space. We assess the interplay between these searches, the standard searches for new scalars decaying directly into SM particles, 
and the measurement of 125 GeV Higgs signal strengths in constraining 2HDM scenarios.
Furthermore, being at the onset of LHC Run 2, we go on to outline the upcoming prospects for direct searches of the neutral scalars $H_0$, $A_0$ 
in the hierarchical 2HDM at the 13 TeV Run of LHC, through the discussion of benchmark plane scenarios. 

After a review of the 2HDM in Section~\ref{sec:2hdm}, we discuss the measurements of Higgs signal strengths in the context of the 
2HDM in Section \ref{sec:HSs}. We then demonstrate the impact of the mass spectrum on LHC searches for $A_0/H_0$ in Sections \ref{A0section} and \ref{H0section} 
as well as briefly commenting on $H^{\pm}$ searches in Section \ref{Hpsection}. In Section \ref{sec:bsm3} we go on to analyze the constraints that can be derived from 
the recent dedicated search of 2HDM neutral scalars with a sizable splitting by the CMS Collaboration~\cite{CMS:2015mba}, highlighting the strong complementarity with the 
Standard searches and analyzing the interplay between these and Higgs measurements discussed in Section \ref{sec:HSs}. 
Finally, in Section \ref{sec:bsmRun2} we present benchmark plane scenarios for searches of these new scalars at LHC Run 2.

\vspace{-3mm}

\section{A (Brief) Review of the 2HDM}
\label{sec:2hdm}

\vspace{-3mm}

In this section we discuss the aspects of the 2HDM relevant to our analysis, defining at the same time 
our notation. We consider a general 2HDM scalar potential with a softly broken $\mathbb{Z}_2$ symmetry in the absence of Charge-Parity (CP) violation, which reads  
\begin{eqnarray}	
\label{2HDM_potential}
V(H_1,H_2) &= &\mu^2_1 \left|H_1\right|^2 + \mu^2_2\left|H_2\right|^2 - \mu^2\left[H_1^{\dagger}H_2+\mathrm{h.c.}\right] \nonumber \\
&+&\frac{\lambda_1}{2}\left|H_1\right|^4 +\frac{\lambda_2}{2}\left|H_2\right|^4 + \lambda_3 \left|H_1\right|^2\left|H_2\right|^2 \\
&+&\lambda_4 \left|H_1^{\dagger}H_2\right|^2+ \frac{\lambda_5}{2}\left[\left(H_1^{\dagger}H_2\right)^2+\mathrm{h.c.}\right]\, ,\nonumber 
\end{eqnarray}
where the two scalar $SU(2)_L$ doublets $H_j$ ($j = 1,2$) may be written as 
\begin{equation}
H_j = \left(\phi_j^{+} , (v_j + h_j + i\,\eta_j)/\sqrt{2} \right)^T\, .
\end{equation}
In addition to the 125 GeV Higgs state $h$, the scalar sector of a 2HDM includes another neutral CP-even scalar $H_0$, a neutral CP-odd scalar $A_0$ and a 
charged scalar $H^{\pm}$. For most of this work, we assume that these new states are heavier than $h$ (it is however possible,
although more experimentally constrained, for either $H_0$ or $A_0$ to be lighter than $m_h =125$ GeV, a possibility which has been explored recently 
in~\cite{Bernon:2014nxa,Bernon:2015wef}). Apart from $m_h$ and $v = 246$ GeV, the scalar potential (\ref{2HDM_potential}) may be parametrized in terms of 
the scalar masses $m_{H_0}$, $m_{A_0}$, $m_{H^{\pm}}$, the squared mass scale $\mu^2$ and two angles $\beta$ and $\alpha$,
the former being related to the ratio of vacuum expectation values ({\it vev}$\,$s) of the two scalar doublets, $v_{1,2}$, via $\mathrm{tan}\,\beta \equiv v_2/v_1$ 
(with $v^2_1 + v^2_2 = v^2$) and the latter parametrising the mixing between the CP-even states. The relation between the physical states $h, \,H_0, \,A_0, \,H^{\pm}$ and 
the states $h_j,\, \eta_j,\, \phi_j^{\pm}$ is given by 
\begin{eqnarray}
\label{rotation_states}
H^{\pm}=-s_{\beta}\, \phi_1^{\pm} + c_{\beta}\, \phi_2^{\pm} & \hspace{9mm} & A_0=-s_{\beta}\, \eta_1 + c_{\beta}\, \eta_2 \nonumber \\
h=-s_{\alpha}\, h_1 + c_{\alpha}\, h_2  & \hspace{9mm} & H_0=-c_{\alpha}\, h_1 - s_{\alpha}\, h_2 \nonumber
\end{eqnarray}
%
with $s_{\beta},c_{\beta}, s_{\alpha},c_{\alpha} \equiv \mathrm{sin}\,\beta, \mathrm{cos}\,\beta, \mathrm{sin}\,\alpha, \mathrm{cos}\,\alpha$, respectively. 
Regarding the couplings of the two doublets $H_{1,2}$ to fermions, the $\mathbb{Z}_2$ in (\ref{2HDM_potential}), even when softly broken by $\mu^2$, may be used to
forbid potentially dangerous tree-level flavour changing neutral currents (FCNCs) by requiring that each fermion type couple to 
one doublet only~\cite{Glashow:1976nt}. By convention, up-type quarks couple to $H_{2}$. In Type I 2HDM all the other fermions 
also couple to $H_{2}$, while for Type II down-type quarks and leptons couple to $H_{1}$. There are two more possibilities (depending on the 
$\mathbb{Z}_2$ parity assignment for leptons with respect to down-type quarks), but we focus here on Types I and II, as they encode the relevant physics of 2HDMs with no tree-level FCNCs. 

The parameters $t_{\beta} \equiv \mathrm{tan}\,\beta$ and $c_{\beta -\alpha} \equiv \mathrm{cos}\,(\beta-\alpha)$ control the strength of the couplings 
of $h$, $H_0$, $A_0$ and $H^{\pm}$ to gauge bosons and fermions. Focusing on the neutral scalars, we denote the couplings normalized to the SM 
values (of $h_{\mathrm{SM}}$) by $\kappa$-factors ($\kappa_V$ for gauge bosons, $\kappa_u$ for up-type quarks, $\kappa_d$ for 
down-type quarks, $\kappa_{\ell}$ for charged leptons), which read 
%
%
 \begin{equation}
 \label{t:kappaI}
 \mathrm{Type}-\mathrm{I}:\,\left\lbrace
 \begin{array}{l}
 \kappa^h_V = s_{\beta-\alpha} \\
 \kappa^h_u = \kappa^h_d = \kappa^h_{\ell} = t^{-1}_{\beta}c_{\beta-\alpha} + s_{\beta-\alpha}  \\
 \kappa^{H_0}_V = - c_{\beta-\alpha} \\
 \kappa^{H_0}_u = \kappa^{H_0}_d = \kappa^{H_0}_{\ell} = t^{-1}_{\beta}s_{\beta-\alpha} - c_{\beta-\alpha}  \\
 \kappa^{A_0}_u = - \kappa^{A_0}_d = - \kappa^{A_0}_{\ell} = t^{-1}_{\beta}
 \end{array}
 \right.
 \end{equation}
  \begin{equation}
 \label{t:kappaII}
 \mathrm{Type}-\mathrm{II}:\,\left\lbrace
 \begin{array}{l}
 \kappa^h_V = s_{\beta-\alpha} \\
 \kappa^h_u = t^{-1}_{\beta}c_{\beta-\alpha} + s_{\beta-\alpha}  \\
 \kappa^h_d = \kappa^h_{\ell} = s_{\beta-\alpha} - t_{\beta}\,c_{\beta-\alpha}\\
 \kappa^{H_0}_V = - c_{\beta-\alpha} \\
 \kappa^{H_0}_u = t^{-1}_{\beta}s_{\beta-\alpha} - c_{\beta-\alpha}  \\
 \kappa^{H_0}_d = \kappa^{H_0}_{\ell} = -t_{\beta}\,s_{\beta-\alpha} - c_{\beta-\alpha}  \\
 \kappa^{A_0}_u = t^{-1}_{\beta}\\
 \kappa^{A_0}_d = \kappa^{A_0}_{\ell} = t_{\beta}
 \end{array}
 \right.
 \end{equation}
For $c_{\beta -\alpha} \to 0$, commonly referred to as the 2HDM {\it alignment} limit, $h$ has SM-like couplings to gauge bosons and fermions ($\kappa^h_i \to 1$, yielding 
$h \to h_{\mathrm{SM}}$), while the coupling $H_0VV$ of $H_0$ to gauge bosons $V = W^{\pm},Z$ vanishes ($\kappa^{H_0}_V \to 0$).

\vspace{2mm}

In order to obtain a viable 2HDM scenario, theoretical constraints from unitarity, perturbativity and stability/boundedness from below of the scalar potential 
(\ref{2HDM_potential}) need to be satisfied. Tree-level stability of the potential $V(H_1,H_2)$ requires $\lambda_1~>~0$, $\lambda_2~>~0$, 
$\lambda_3~>~-\sqrt{\lambda_1\lambda_2}$, $\lambda_3+\lambda_4-|\lambda_5|>- \sqrt{\lambda_1\lambda_2}$ (see {\it e.g.}~\cite{Maniatis:2006fs}).
At the same time, tree-level unitarity\footnote{For a recent one-loop analysis, leading to slightly more stringent bounds, see~\cite{Grinstein:2015rtl}.} 
imposes bounds on the size of various combinations of the quartic couplings $\lambda_i$~\cite{Akeroyd:2000wc}, 
like $\left|\lambda_3 \pm \lambda_4 \right| < 8 \pi$, $\left|\lambda_3 \pm \lambda_5 \right| < 8 \pi$, $\left|\lambda_3 + 2 \lambda_4 \pm 3 \lambda_5 \right| < 8 \pi$ and 
{\small$\left|\lambda_1 + \lambda_2 \pm \sqrt{(\lambda_1-\lambda_2)^2 + 4 \lambda_4^2} \right|$} $< 16 \pi$.~Similar (although generically less stringent) 
bounds on $\lambda_i$ may be obtained from perturbativity arguments.
We may express $\lambda_i$ in terms of the physical scalar masses, the mixing angles $\alpha$, $\beta$ and $\mu^2$:
\bea
\label{couplings1}
\lambda_1&=&\frac{1}{v^2\, c^2_{\beta}} \left(-\mu^2\,t_{\beta} + m^2_h \, s^2_{\alpha} + m^2_{H_0} \, c^2_{\alpha}\right),\\
\label{couplings2}
\lambda_2&=&\frac{1}{v^2\, s^2_{\beta}} \left(-\mu^2\,t^{-1}_{\beta} + m^2_h \, c^2_{\alpha} + m^2_{H_0} \, s^2_{\alpha}\right),\\
\label{couplings3}
\lambda_3&=&\frac{1}{v^2} \Big[-\frac{2\mu^2}{s_{2\beta}}  + 2 m_{H^\pm}^2 + \left(m_{H^0}^2-m_{h}^2\right)\frac{s_{2\alpha}}{s_{2\beta}}\Big],\\
\label{couplings4}
\lambda_4&=&\frac{1}{v^2} \left(\frac{2\mu^2}{s_{2\beta}}+m_{A^0}^2-2 m_{H^\pm}^2\right),\\
\label{couplings5}
\lambda_5&=&\frac{1}{v^2}\left(\frac{2\mu^2}{s_{2\beta}} - m_{A^0}^2\right).
\eea
As seen from (\ref{couplings1}-\ref{couplings5}), for a given set of values for 
$m_{H_0}$, $m_{A_0}$, $m_{H^{\pm}}$, $t_{\beta}$ and $c_{\beta -\alpha}$, only a certain range for $\mu^2$ is allowed by the 
combination of these theoretical constraints. 
In particular, $\lambda_{1,2} > 0$ directly imply an upper bound on $\mu^2$ from (\ref{couplings1})-(\ref{couplings2}). 
It is however possible that no value of $\mu^2$ allows to satisfy all three theoretical requirements simultaneously, in which case such a   
set of values for the scalar masses and mixing angles would not be viable. 
If an allowed $\mu^2$ range exists, the size of trilinear scalar couplings such as $\lambda_{H_0 h h}$ and $\lambda_{H_0 A_0 A_0}$ 
(which control the partial widths $\Gamma_{H_0 \to h h}$, $\Gamma_{H_0 \to A_0 A_0}$ when these decays are kinematically allowed) 
or $\lambda_{h H^{+} H^{-}}$ (which controls the size of the charged scalar loop contribution to the $h \to \gamma\gamma$ decay amplitude, given by 
$\Delta^{\pm}_{\gamma}$) depend on the value of $\mu^2$. Indeed, the trilinear couplings $\lambda_{H_0 h h}$ and $\lambda_{H_0 A_0 A_0}$ are given by
{\small
\begin{align}
    \label{lambdaHhh}
    \hspace{-2.2mm} v\lambda_{H_0 h h} &= \frac{2\,c_{\beta -\alpha}}{s_{2\beta}}\Bigg[\left(1 -
     3\frac{s_{2\alpha}}{s_{2\beta}}\right)\mu^2
    \hfill+ (2 m^2_h + m^2_{H_0})\frac{s_{2\alpha}}{2} \Bigg]
\end{align}}
{\small
\bea
\label{lambdaHAA}
v\lambda_{H_0 A_0 A_0} &=& 2\Big[ \cba(2 m^2_{A_0}+m^2_{H_0})  
\\
&-& 2\left(\sba\frac{c_{2\beta}}{s_{2\beta}} - \cba\right)\left(m_{H_{0}}^2 - \frac{\mu^2}{\,s_\beta\,c_\beta}\right) \Big] \nonumber
\eea
}
\noindent Apart from vanishing in the alignment limit, if $s_{2\beta} - 3 s_{2\alpha} \neq 0$, the coupling $\lambda_{H_0 h h}$ also vanishes for 
$\mu^2 = (2 m^2_h + m^2_{H_0}) (s_{2\alpha}s_{2\beta})/(6s_{2\alpha}-2s_{2\beta})$, if such value of $\mu^2$ lies within 
the allowed range.
Similarly, in the alignment limit $\lambda_{H_0 A_0 A_0}$ vanishes for $t_{\beta} = 1$ or $\mu^2 = m^2_{H_0}\,s_\beta\,c_\beta$. 
The trilinear coupling $\lambda_{h H^{+} H^{-}}$ reads
%
%
%
{\small
\bea
\label{lambdahHpHm}
v\lambda_{hH^+H^-} &=&  \Big[ \sba\left(m_h^2-2m_{H^\pm}^2\right)  
\\
&-& 2\left(\cba \frac{c_{2\beta}}{s_{2\beta}} +\sba\right)\left(m_h^2 - \frac{\mu^2}{s_\beta\,c_\beta}\right) \Big], \nonumber
\eea
}
%

\noindent so that $\Delta^{\pm}_{\gamma}$ inherits a dependence on $\mu^2$ and other 2HDM parameters besides $m^2_{H^{\pm}}$ through 
$\lambda_{h H^{+} H^{-}}$. These trilinear couplings illustrate the phenomenological impact of the soft $\mathbb{Z}_2$-breaking parameter in the 2HDM, 
which will be analyzed in more detail in Section \ref{sec:3}.

\vspace{-3mm}

\section{Hierarchical {\it vs} Degenerate 2HDM: The LHC Run 1 Legacy}
\label{sec:3}

\vspace{-3mm}

Let us now concentrate on the mass spectrum of the 2HDM. We first note that constraints from measurements of EW precision observables (EWPO), in particular of the 
$T$-parameter, generically require $H^{\pm}$ to be relatively degenerate with either $A_0$ or $H_0$~\cite{Grimus:2007if}. From a phenomenological 
perspective we can then distinguish between a {\it degenerate} spectrum where all mass splittings among the new scalar states are small,
$|m_{A_0} - m_{H_0}| \ll m_Z$, and a {\it hierarchical} spectrum for which the mass splitting among the new neutral scalars is sizable, $|m_{A_0} - m_{H_0}|  \gtrsim m_Z$.

The main phenomenological feature of a hierarchical 2HDM spectrum is that the decays $\varphi_i \to \varphi_j V$, with $\varphi_{i,j} = H_0, \,A_0, \,H^{\pm}$ ($i\neq j$) 
and $V = W^{\pm}, Z$ become kinematically allowed and generically yield the dominant branching fraction, with the decays into SM states comparatively suppressed. 
These considerations motivate performing a comparison of the allowed 2HDM parameter space for both types of spectra, assessing the impact of sizeable mass splitting(s). 
In this respect, key probes of 2HDM scenarios are ATLAS/CMS measurements of Higgs signal strengths and searches 
for new scalar states at the LHC. 


\vspace{-3mm}

\subsection{Higgs Signal Strengths in the 2HDM}
\label{sec:HSs}

\vspace{-3mm}

The values for the Higgs signal strengths measured by the ATLAS and CMS experiments during the 7 and 8 TeV LHC runs 
set an important constraint on the 2HDM parameter space~\cite{Aad:2015pla} 
(see also~\cite{Celis:2013rcs,Krawczyk:2013gia,Grinstein:2013npa,Chen:2013rba,Craig:2013hca,Eberhardt:2013uba,Dumont:2014wha,Bernon:2014vta,Craig:2015jba}). 
The model prediction for the signal strength in a final state $xx$ is given by \mbox{$\mu^{\mathrm{2HDM}}_{xx} = \sum_i \epsilon_i \times \mu^i_{xx}$}, 
with $\epsilon_i$ corresponding to the relative
contribution to the signal from a particular Higgs production mode $i$, and $\mu^i_{xx}$ being the 2HDM signal strength for that production mode
\be
\label{SignalStrength_2HDM}
\small{\mu^i_{xx} = 
\frac{\left[ \sigma_i(p p \to h) \times \mathrm{BR}(h \to xx) \right]_{\mathrm{2HDM}}}{\left[  \sigma_i(p p \to h) \times \mathrm{BR}(h \to xx)\right]_{\mathrm{SM}}}}\, ,
\ee
to be compared with the values obtained by ATLAS and CMS analyses in the relevant detection channels, namely 
$h \to W W^*$~\cite{ATLAS:2013wla,Aad:2013wqa,TheATLAScollaboration:2013hia,Chatrchyan:2013iaa,CMS:2013xda}, 
$h \to Z Z^*$~\cite{Aad:2013wqa,Aad:2014eva,Chatrchyan:2013mxa}, 
$h \to \gamma \gamma$~\cite{Aad:2014eha,Khachatryan:2014ira}, $h \to \bar{b}\,b$~\cite{TheATLAScollaboration:2013lia,Chatrchyan:2013zna} 
and $h \to \tau\,\tau$~\cite{ATLAS_tau,Chatrchyan:2014nva}. 
In all but one of these channels the various $\mu^i_{xx}$ are directly obtained from the $\kappa^h_x$ factors 
in (\ref{t:kappaI}) and (\ref{t:kappaII}), depending only on $c_{\beta-\alpha}$ and $t_{\beta}$. The sole exception is $\mu^i_{\gamma\gamma}$, 
since BR($h\to \gamma\gamma$) also involves the contribution to the $h \to \gamma\gamma$ decay amplitude from the charged scalar loop, $\Delta^{\pm}_{\gamma}$, 
which introduces a dependence on $\mu^2$ and other physical parameters via the trilinear $\lambda_{hH^+H^-}$. As a result, a comparison to the 
experimental data would strictly speaking require a generalized $\Delta\chi^2$ likelihood fit in a multidimensional parameter space subject to the theoretical 
constraints on $\mu^2$ above discussed. 
However, since the charged scalar loop generically gives a very subdominant contribution, 
we adopt here a simplified approach of neglecting this term by setting $\lambda_{hH^+H^-}=0$.
The Higgs signals constraints can then be obtained by performing a $\Delta\chi^2$ likelihood fit to the 2HDM parameters $c_{\beta-\alpha}$ and $t_{\beta}$, for 
which we use the public codes {\sc Lilith}~\cite{Bernon:2015hsa} and {\sc HiggsSignals}~\cite{Bechtle:2013xfa,Bechtle:2014ewa}.
 The values of $\epsilon_i$ in (\ref{SignalStrength_2HDM}) may be obtained from the experimental analyses and are provided in both these programs
({\it e.g.} for {\sc HiggsSignals} they may be found in Appendix A of~\cite{Bechtle:2014ewa}). 


The results are shown in Figure \ref{fig:HiggsFit} for Type I (\textsl{Left}) and Type II (\textsl{Right}) 2HDM. 
The green areas correspond to the 95\% C.L. allowed region from {\sc Lilith}, while the hatched-purple ones are those from 
{\sc HiggsSignals}. Both show good agreement with the ATLAS experimental fit~\cite{Aad:2015pla}, the fit 
from {\sc Lilith} being slightly more constraining than both {\sc HiggsSignals} and ATLAS. 
In Type I, a sizable departure from alignment is allowed as soon as $t_{\beta} \gtrsim 1$, and 
the limit on $c_{\beta-\alpha}$ becomes both independent of $t_{\beta}$ and symmetric around $c_{\beta-\alpha} = 0$ for $t_{\beta} \gg 1$, which 
can be understood from (\ref{t:kappaI}). For Type II, there are two distinct allowed regions: {\it (i)} the region close to the 
alignment limit $c_{\beta-\alpha} \ll 1$ corresponding to a SM-like Higgs $h$, 
with a mild preference for $c_{\beta-\alpha} > 0$ and $t_{\beta} \sim 1$; {\it(ii)} the {\it wrong-sign} scenario 
$s_{\beta+\alpha} \sim 1$, for which $\kappa^h_d < 0$, and $0 < 1 + \kappa^h_d \ll 1$ (see {\it e.g.}~\cite{Ferreira:2014naa} for a detailed discussion of this limit, 
possible only in Type II).

\begin{widetext}
\onecolumngrid
\begin{figure}[t!]
\begin{center}
\includegraphics[width=0.85\textwidth]{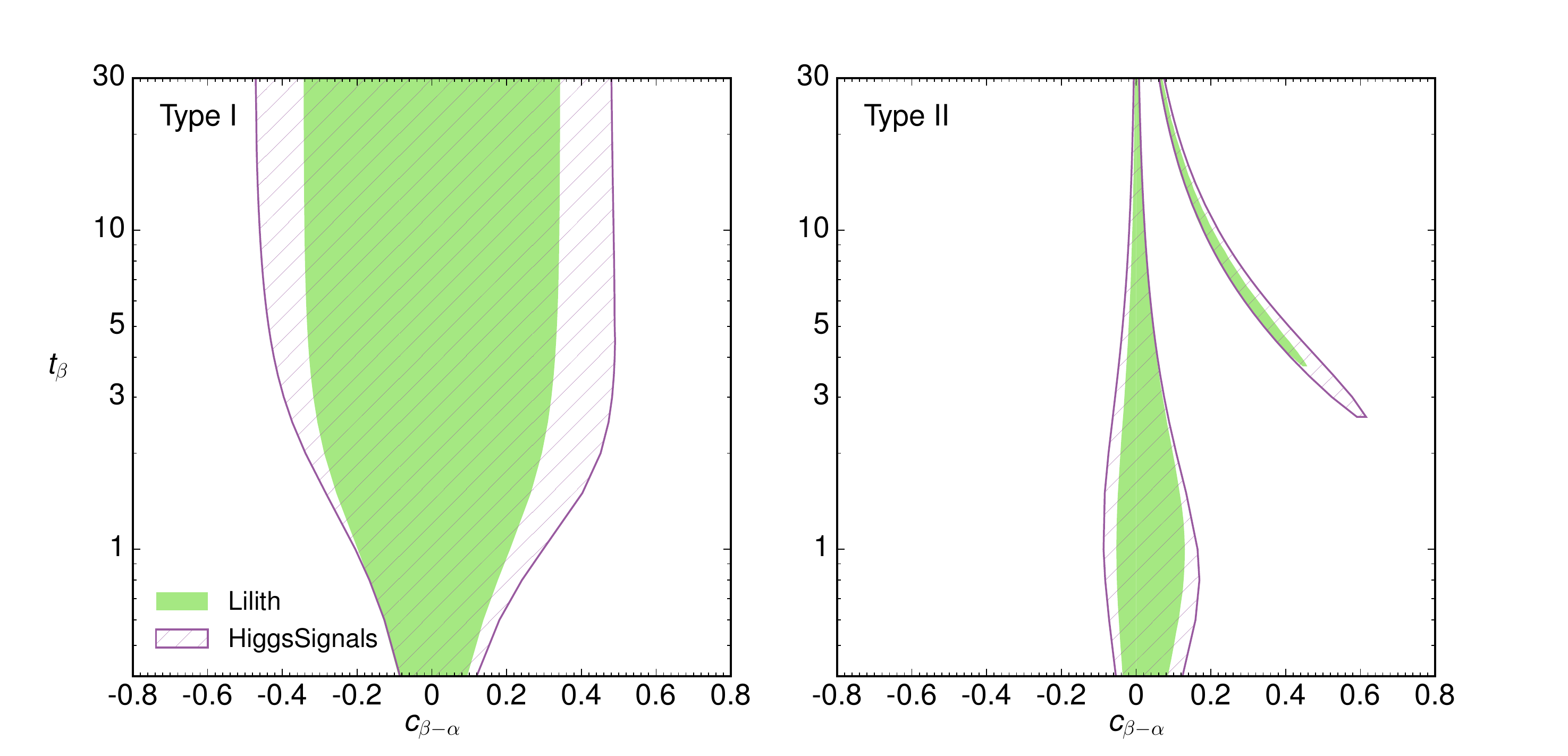}
\caption{\small 95\% C.L. Likelihood fit to Higgs signal strengths in the ($c_{\beta-\alpha}, t_{\beta}$) plane, for Type I (\textsl{Left}) and Type II (\textsl{Right}) 2HDM,
using {\sc Lilith} (solid-green region) and {\sc HiggsSignals} (hatched-purple region). See text for details on the fit.}
\vspace{-10mm}
\label{fig:HiggsFit}
\end{center}
\end{figure}
\end{widetext}

\subsection{LHC Searches for $A_0$ into SM States}
\label{A0section}

\vspace{-3mm}

We discuss now the limits on the 2HDM parameter space from ATLAS and CMS searches of $A_0$ decaying via 
$A_0 \to Z h$ ($h\to \bar{b} b, h \to \tau\tau$)~\cite{Aad:2015wra, Khachatryan:2015lba}, via $A_0 \to \gamma \gamma$~\cite{Aad:2014ioa,CMS:2014onr}
and $A_0 \to \tau \tau$~\cite{Aad:2014vgg,Khachatryan:2014wca}. 
For the $A_0 \to \tau \tau$ searches, the production of $A_0$ in association with a $\bar{b} b$ pair is taken into account by the ATLAS/CMS experimental 
analyses in addition to production through gluon fusion, the former being important for Type II at large values of $t_{\beta}$. 
Furthermore, we stress that while the search via $A_0 \to Z h$ vetoes any b-tagged jets beyond those from $h\to \bar{b} b$ (see {\it e.g.}~\cite{Aad:2015wra}), the 
$b$-jets resulting from the $ p p \to \bar{b} b \,A_0$ process generically have large rapidity values and consequently yield a very low b-tagging efficiency~\cite{CMS:2013xfa}. 
Thus we also consider $\bar{b} b$-associated production of $A_0$ in the $A_0 \to Z h$ searches, and do not implement a b-tagging efficiency 
suppression in this case. 

In order to derive the bounds on the 2HDM parameter space, we compute the $A_0$ production cross-section in gluon fusion and 
in association with $\bar{b} b$ at NNLO in QCD with {\sc SusHi}~\cite{Harlander:2012pb}) for Types I and II as a function of $t_{\beta}$ and $m_{A_0}$,  
and then use {\sc 2HDMC}~\cite{Eriksson:2009ws} to compute the branching fractions for $A_0 \to \tau \tau$, $A_0 \to \gamma \gamma$, 
$A_0 \to Z h$ and $h \to \bar{b}b,\tau\tau$ as a function of $t_{\beta}$, $c_{\beta-\alpha}$, $m_{A_0}$ and $m_{H_0}$.
The $95 \,\%$ C.L. exclusion region in the ($c_{\beta-\alpha}$, $t_{\beta}$) plane resulting from these searches is shown 
in Figure \ref{fig:1} for different values of $m_{A_0}$ and $m_{A_0} - m_{H_0}$, and discussed below.

Let us consider first a high mass scenario for $A_0$, above the $\bar{t} t$ threshold: The exclusion region for $m_{A_0} = 500$~GeV in Types I 
and II is shown respectively in Figure \ref{fig:1} (Top-\textsl{Left}) and (Top-\textsl{Right}).
The only sensitive channel above the $\bar{t} t$ threshold is $A_0 \to Z h$, and for Type II also $A_0 \to \tau \tau$ in $\bar{b} b$-associated production.
Nevertheless, we see that for low/moderate $t_{\beta}$, these searches only constrain values of $\left|c_{\beta-\alpha} \right| \gtrsim 0.15$.
The green region corresponds to the exclusion for $m_{H_0} = 500$~GeV, when $A_0$ can only decay into SM states.
As $m_{H_0}$ decreases and the decay $A_0 \to H_0 Z$ becomes kinematically allowed, the current limits from searches of SM decay channels weaken significantly, as 
the orange and purple regions in Figure \ref{fig:1} (Top) show respectively for $m_{H_0} = 300$ GeV and $m_{H_0} = 150$ GeV. 

The impact of a sizable $m_{A_0} - m_{H_0}$ splitting is even more important for $m_{A_0}$ below the $\bar{t} t$ threshold:
The excluded region for $m_{A_0} = 300$~GeV is shown in Figure \ref{fig:1} (Medium) for Type I (\textsl{Left}) and Type II (\textsl{Right}), 
both in the degenerate scenario $m_{H_0} = 300$~GeV (green region) and for a hierarchical scenario with $m_{H_0} = 150$~GeV (purple region).
In the former, the limits from $A_0 \to Z h$ searches are stringent, ruling out $\left|c_{\beta-\alpha} \right| \gtrsim 0.02$ 
for $t_{\beta} < 6$ in Type I. Even for $c_{\beta-\alpha} \to 0$, $A_0 \to \gamma\gamma$ and $A_0 \to \tau \tau$ searches constrain the region of 
$t_{\beta} \lesssim 2$ and $t_{\beta} \lesssim 3$ respectively for Types I and II. In contrast, for the hierarchical scenario the 
$A_0 \to \gamma\gamma$ and $A_0 \to \tau \tau$ searches only constrain values of $t_{\beta} \lesssim 0.5$, 
 while the sensitivity of the $A_0 \to Z h$ searches also reduces drastically.

Finally, we also present the limits for a light $A_0$, with $m_{A_0} = 150$~GeV in Figure \ref{fig:1} (Bottom). In this case we do not consider a hierarchical 2HDM scenario
(with $A_0$ being the heavier state), as it would require $c_{\beta-\alpha} \to 0$ to avoid non-observation of $H_0$ at LEP (we will however briefly discuss 
this region of parameter space in Section \ref{sec:bsm3}).
Both for Type I (\textsl{Left}) and Type II (\textsl{Right}), the $A_0 \to \tau \tau$ and $A_0 \to \gamma \gamma$ searches yield the constraint $t_{\beta} \gtrsim 1.5$, 
while for Type II the searches for $A_0 \to \tau \tau$ in $\bar{b} b$-associated production also yield a limit $t_{\beta} < 10$.

\begin{widetext}
\onecolumngrid
\begin{figure}[h!]
\begin{center}
\includegraphics[width=0.9\textwidth]{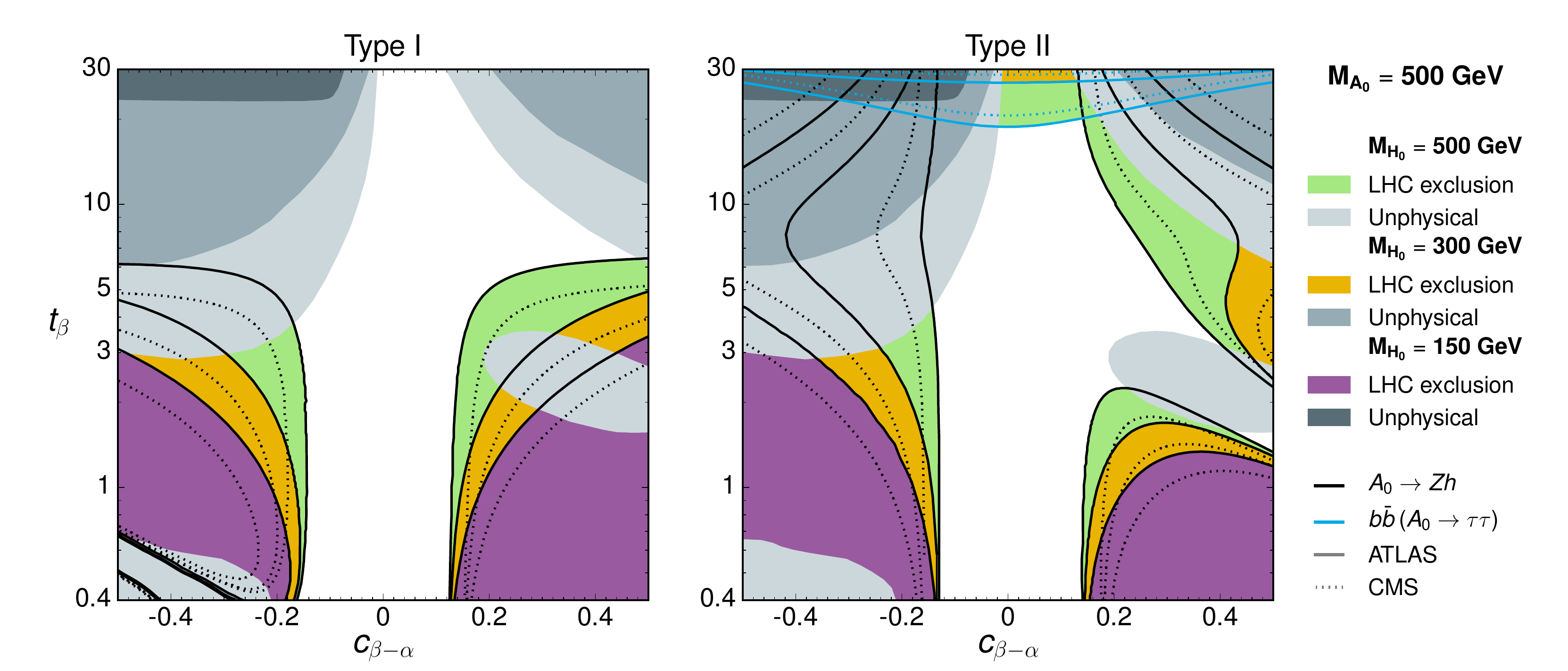}

\vspace{2mm}
\includegraphics[width=0.9\textwidth]{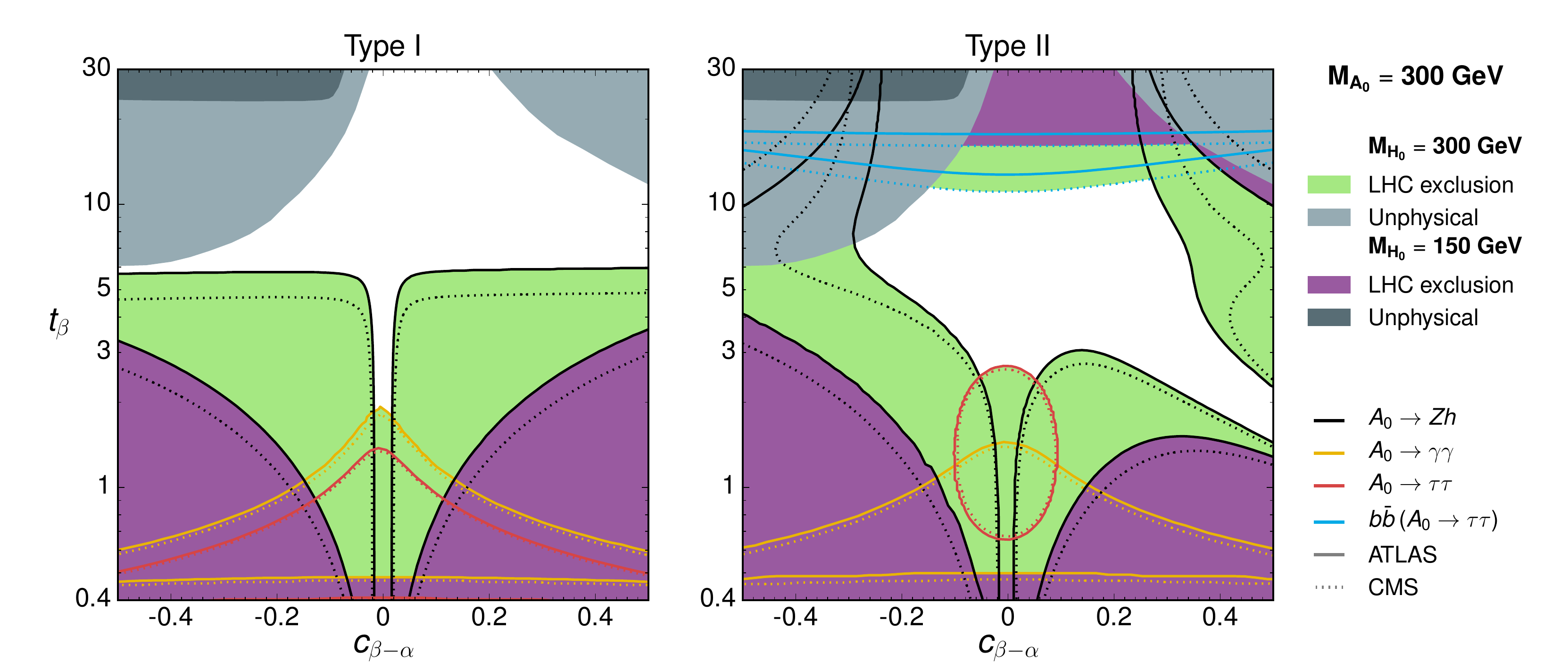}

\vspace{2mm}
\includegraphics[width=0.9\textwidth]{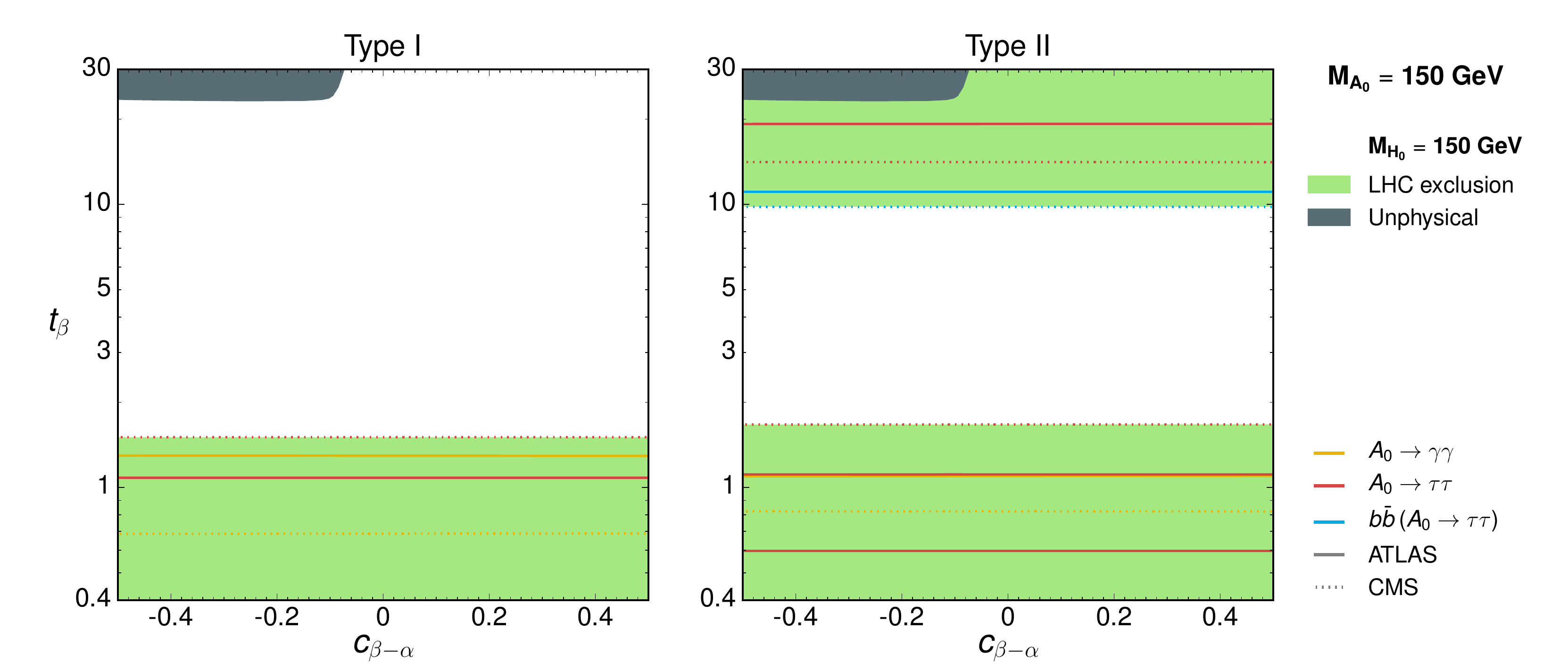}
\caption{\small Current $95 \,\%$ C.L. excluded region by ATLAS (solid lines) and CMS (dashed lines), respectively for $m_{A_0} = 500,\,300,\,150$~GeV (Top/Medium/Bottom) 
and for 2HDM Type I/II (\textsl{Left}/\textsl{Right}), coming from searches of $g g, b \bar{b} \to A_0 \to Z h\,\, (h\to b \bar{b})$ (black lines), 
$g g \to A_0 \to \gamma \gamma$ (yellow lines), $g g \to A_0 \to \tau \tau$ (red lines) and $b \bar{b} \to A_0 \to \tau \tau$ (light-blue lines).
In each case, the limits in the degenerate scenario $m_{H_0} = m_{A_0}$ are shown in green, while those for the hierarchical scenario(s) are shown 
in purple/orange. The various grey regions correspond to the theoretically excluded regions for the degenerate and hierarchical 2HDM scenarios (darker grey as 
$m_{H_0}$ is lower).}
\label{fig:1}
\end{center}
\end{figure}
\end{widetext}

The above discussion highlights the fact that, in the presence of a sizable mass splitting $m_{A_0} - m_{H_0}$, searches for $A_0$ decaying into SM final states 
have little sensitivity, and in particular do not yield further information on the allowed 2HDM parameter space to that obtained from Higgs coupling measurements, 
discussed in the previous Section. Let us however emphasize that the lighter state in the hierarchical 2HDM scenario, in this case $H_0$,
would decay solely into SM states. Thus, the constraints on the parameter space from searches of $H_0$ into SM final states would fully apply for a hierarchical
scenario with $m_{A_0} - m_{H_0} > 0$. LHC searches for $H_0$ will be analyzed in Section \ref{H0section}. 

Before moving on to the next Section, let us discuss the impact of theoretical constraints from unitarity, perturbativity and stability of the 2HDM scalar potential 
on Figure \ref{fig:1}, where the theoretically excluded regions are shown in grey. Focusing on the case $m_{H^{\pm}} = m_{A_0}$, 
and defining $m^2_{A_0} - m^2_{H_0} \equiv \Delta^2 \geq 0$, 
Figure \ref{fig:1} shows that the exclusion becomes more important as $m_{H_0}$ increases, particularly for $t_{\beta} \gg 1$. 
The departure from alignment also has a strong impact on the theoretically viable parameter space, specially for $c_{\beta-\alpha} < 0$. These features may be understood 
from the interplay of $\lambda_1 > 0$ and various unitarity limits. Writing $\lambda_1$ as
%
\bea
\label{lambda1_Stability}
\lambda_{1}v^2& = & m_h^2 - t_{\beta}(1+t_{\beta}^2)\, \Omega^2  \nonumber \\
& - & (m^2_{H_0} - m_h^2) \left[c_{\beta-\alpha}^2 (t_{\beta}^2-1) - 2 t_{\beta} s_{\beta-\alpha} c_{\beta-\alpha}  \right] \hspace{4mm}
\eea
%
\noindent with $\Omega^2 \equiv \mu^2 - m^2_{H_0} s_{\beta} c_{\beta}$, we see that for $m^2_{H_0} \gg m^2_h$ (neglecting $m_h^2$ in (\ref{lambda1_Stability})) 
and $t_{\beta} > 1$, $\Omega^2 < 0$ is required to satisfy $\lambda_1 > 0$ for either $c_{\beta-\alpha} < 0$ or $c_{\beta-\alpha} t_{\beta} \gg 1$.
This in turn impacts the unitarity requirements, {\it e.g.}
{\small 
\bea
\label{Unitarity}
\left|\lambda_3 + \lambda_4 \right| \sim  \left|\frac{\Delta^2}{v^2}  + \frac{m^2_{H_0}c_{\beta-\alpha}}{v^2}    \left[s_{\beta-\alpha} (t_{\beta} - t_{\beta}^{-1}) 
- 2 c_{\beta-\alpha}\right]\right| < 8 \pi \nonumber \\
\left|\lambda_3 + 2\lambda_4 + 3\lambda_5\right|  \sim  \left|-\frac{3\Delta^2}{v^2}  + \frac{4}{s_{\beta} c_{\beta}} \frac{\Omega^2}{v^2} \right. \hspace{3.25cm} \nonumber \\
\left. + \frac{m^2_{H_0}}{v^2}   c_{\beta-\alpha} \left[s_{\beta-\alpha} (t_{\beta} - t_{\beta}^{-1}) 
- 2 c_{\beta-\alpha}\right]\right| < 8 \pi \hspace{8mm}
\eea
}
which are then violated for $t_{\beta} \gg 1$ and/or $m^2_{H_0} \gg v^2$, as no cancellation among terms is possible in both 
$\left|\lambda_3 + \lambda_4 \right|$ and $\left|\lambda_3 + 2\lambda_4 + 3\lambda_5\right|$ (since $\Delta^2 \geq 0$, $\Omega^2 < 0$). 

We note that the above requirement $\Omega^2 < 0$ to satisfy $\lambda_1 > 0$ may be avoided for 
$c_{\beta-\alpha} (t_{\beta}^2-1) - 2 t_{\beta} s_{\beta-\alpha} \sim 0$, for which the last term 
in (\ref{lambda1_Stability}) vanishes. This cancellation, which happens for $c_{\beta-\alpha} = 2 t_{\beta}/(1+t^2_{\beta})$, is observed for $m^2_{H_0} = 500$ GeV and 
$c_{\beta - \alpha} > 0$ in Figure  \ref{fig:1} (Top). We also note that in exact alignment $c_{\beta - \alpha} = 0$, $\Omega^2 = 0$ automatically yields $\lambda_1 > 0$ 
(and all other boundedness-from-below requirements are also trivially satisfied for $\Delta^2 \geq 0$). The unitarity constraints are then only violated for 
$\Delta^2 \gg v^2$, and thus $c_{\beta - \alpha} = 0$ is always allowed in Figure \ref{fig:1}.
 

\subsection{LHC Searches for $H_0$ into SM States}
\label{H0section}

\vspace{-3mm}

We turn on now to analyze the constraints from LHC searches for $H_0$. 
The relevant searches to be considered are $H_0 \to ZZ \to \ell \ell \ell \ell$~\cite{ATLAS:2013nma} (and in the low mass region also $H_0 \to WW$~\cite{ATLAS:2014aga}), 
$H_0 \to hh\to \bar{b} b \gamma\gamma$~\cite{Aad:2014yja} by ATLAS and $H_0 \to WW,ZZ$~\cite{Khachatryan:2015cwa} (both low and high mass region),
$H_0 \to hh\to b \bar{b} \gamma\gamma$~\cite{CMS:2014ipa} and $H_0 \to hh\to \bar{b} b \bar{b} b$~\cite{Khachatryan:2015yea} by CMS.
In all these searches, $\bar{b} b$-associated production of $H_0$ is implicitly included\footnote{For $H_0 \to WW,\,ZZ$ searches, 
$\bar{b} b$-associated production generally fails the Vector Boson Fusion and $V$-associated production analysis tags, and so is included in the gluon fusion
category. For $H_0 \to hh$ searches, the analysis is inclusive {\it w.r.t.} $H_0$ production.}
together with gluon fusion.
In addition, the ATLAS/CMS searches via $A_0/H_0 \to \gamma \gamma$~\cite{Aad:2014ioa,CMS:2014onr} and via $A_0/H_0 \to \tau \tau$~\cite{Aad:2014vgg,Khachatryan:2014wca} 
discussed in the previous Section also apply in this case.

As in the previous section, we use {\sc SusHi} to compute the gluon fusion and $\bar{b} b$-associated $H_0$ production cross-sections 
at NNLO in QCD for Types I and II as a function of $c_{\beta-\alpha}$, $t_{\beta}$ and $m_{H_0}$, and then use 
{\sc 2HDMC} to compute the branching fractions for $H_0 \to \tau \tau$, $H_0 \to \gamma \gamma$, 
$H_0 \to Z Z$, $H_0 \to hh$ and $h \to \bar{b}b,\gamma\gamma$ as a function of $c_{\beta-\alpha}$, $t_{\beta}$, $m_{H_0}$, $m_{A_0}$ and $\mu^2$.
We stress that, contrary to the $A_0$ case, the value of $\mu^2$ has a significant impact on the $H_0$ branching fractions via the modification 
of the trilinear coupling $\lambda_{H_0 h h}$, which changes the $H_0 \to hh$ partial width (recall the discussion at the end of Section \ref{sec:2hdm}). 
In order to account for the dependence of $\mu^2$ on the $95 \,\%$ C.L. limits, we compute the theoretically viable $\mu^2$ 
range as a function of $c_{\beta-\alpha}$, $t_{\beta}$, $m_{H_0}$ and $m_{A_0}$, and derive the bounds 
on the values of $\mu^2$ that respectively minimize ($\mu^2_{\mathrm{min}}$) and maximize ($\mu^2_{\mathrm{max}}$) the $H_0 \to hh$ branching 
fraction within the allowed $\mu^2$ range. 

We begin now by discussing the scenario with a light $H_0$, and consider the $95 \,\%$ C.L. exclusion region for 
$m_{H_0} = 150$~GeV in the degenerate scenario, as shown in Figure \ref{fig:2bis}. Due to the absence of the $H_0 \to hh$ decay 
in this case, the $H_0 \to Z Z^*$ and $H_0 \to \tau \tau$ branching fractions are not sensitive to the value of $\mu^2$, and only 
$H_0 \to \gamma\gamma$ is mildly dependent via the $H^{\pm}$ loop contribution. Nevertheless, Figure \ref{fig:2bis} shows that the important limits 
in the $c_{\beta-\alpha}$, $t_{\beta}$ plane are given by $H_0 \to Z Z^*$ and $H_0 \to \tau \tau$ searches, with $H_0 \to \gamma\gamma$ less sensitive.
As has been emphasized in Section \ref{A0section}, the present limits are complementary to those from $A_0$ searches in the hierarchical 2HDM, {\it e.g.}~for the 
($m_{A_0}$, $m_{H_0}$) = ($300$, $150$)~GeV and ($500$, $150$)~GeV benchmarks considered in Figure \ref{fig:1}. 


In Figure \ref{fig:3} we show the limits from $H_0$ searches for $m_{H_0} = 300$~GeV, for Type I/II (\textsl{Left}/\textsl{Right}). Here, the presence of the decay 
$H_0 \to hh$ requires us to take into account the $\mu^2$ dependence in the limit extraction, and we show the limits for $\mu^2 = \mu^2_{\mathrm{min}}$
(Top) and $\mu^2 = \mu^2_{\mathrm{max}}$ (Bottom). In the former case the strongest limits come from $H_0 \to Z Z$ searches, with $H_0 \to h h$  
playing no relevant role because of its suppressed branching fraction. Moreover, in this case the presence of a 
sizable $m_{H_0} - m_{A_0}$ splitting does lead to a significant reduction of the limits on the 2HDM parameter space from these searches. 
In contrast, for $\mu^2 = \mu^2_{\mathrm{max}}$ the $H_0 \to h h$ searches provide the dominant constraint for low and moderate  $t_{\beta}$, and these limits do not change 
significantly in a hierarchical 2HDM scenario, as the branching fraction of $H_0 \to h h$ is still the dominant one in this case. 
A similar situation occurs for $m_{H_0} = 500$~GeV, as shown in Figure \ref{fig:4}. Again, for $\mu^2 = \mu^2_{\mathrm{min}}$
(Top) $H_0 \to Z Z$ searches provide the only meaningful constraint, which gets significantly weakened in the hierarchical scenario 
$m_{H_0} - m_{A_0} \gg m_Z$. 
For $\mu^2 = \mu^2_{\mathrm{max}}$ the LHC searches for $H_0 \to h h$ in $b \bar{b} b \bar{b}$ and 
$b \bar{b} \gamma \gamma$ are the most constraining, being particularly sensitive around $t_{\beta} \sim 1$, and the limits only get mildly weakened 
in the hierarchical 2HDM scenario. In addition, for $m_{H_0} = 500$~GeV there is no appreciable difference between Types I and II for low and 
moderate $t_{\beta}$, with $b\bar{b}$-associated production of $H_0$ in $H_0 \to \tau \tau$ searches constraining the $t_{\beta} \gg 1$ region in Type II.

Contrary to Figure \ref{fig:1}, Figures \ref{fig:3}-\ref{fig:4} do not show the would-be limits on the 
($c_{\beta-\alpha}$, $t_{\beta}$) plane from searches of $H_0$ in regions which are not viable theoretically: 
These limits depend crucially on the value of $\mu^2$ (in contrast with the situation for $A_0$ searches discussed in Section 
\ref{A0section}), and the theoretical bounds correspond precisely to the absence of an allowed $\mu^2$ range.

\begin{widetext}
\onecolumngrid
\begin{figure}[h!]
\begin{center}
\includegraphics[width=0.90\textwidth]{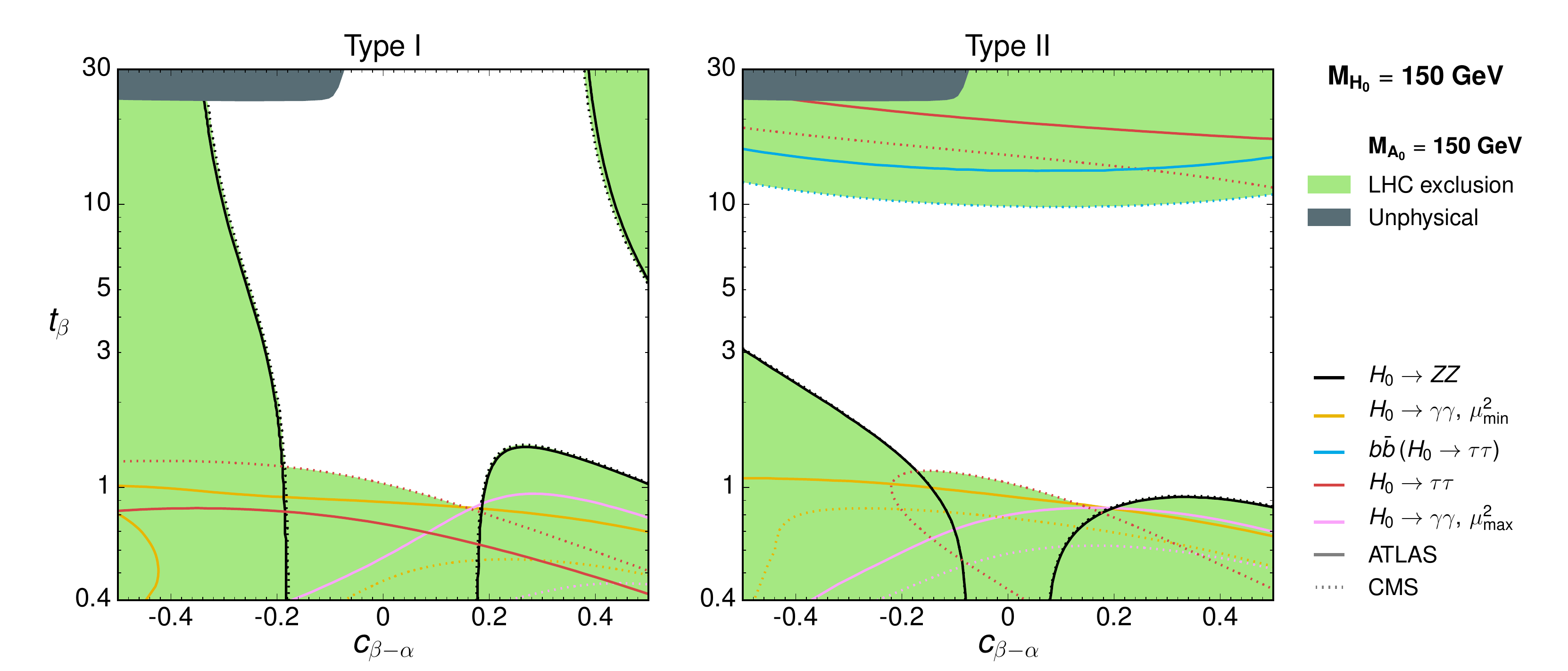}
\caption{\small Current $95 \,\%$ C.L. excluded region (in green) by ATLAS (solid lines) and CMS (dashed lines), respectively for $m_{H_0} = 150$~GeV 
and 2HDM Type I/II (\textsl{Left}/\textsl{Right}), coming from searches of $g g, b \bar{b} \to H_0 \to WW,\,ZZ$ (black lines), $g g \to A_0 \to \tau \tau$ 
(red lines), $b \bar{b} \to A_0 \to \tau \tau$ (light-blue lines) and $g g \to A_0 \to \gamma \gamma$ respectively for $\mu^2 = \mu^2_{\mathrm{min}}$/$\mu^2 = \mu^2_{\mathrm{max}}$ 
(yellow/pink lines). The grey region is theoretically excluded.}
\label{fig:2bis}
\end{center}
\end{figure}
\end{widetext} 

Figures \ref{fig:3}-\ref{fig:4} highlight that, for $\Delta^2 < 0$ (and $m^2_{H^{\pm}} = m^2_{H_0}$), the theoretical bounds from 
stability, unitarity and perturbativity are significantly more important than for the previously discussed $\Delta^2 > 0$ case, 
and in particular constrain the alignment limit $c_{\beta -\alpha} = 0$. The stability conditions for $c_{\beta -\alpha} = 0$ read 
{\small 
\bea
\label{Stability2}
\lambda_1 &=& \frac{m_h^2}{v^2} - t_{\beta}(1+t_{\beta}^2)\, \frac{\Omega^2}{v^2}   > 0 \nonumber \\
\lambda_2  &=& \frac{m_h^2}{v^2} - \frac{(1+t^{-2}_{\beta})}{t_{\beta}}\, \frac{\Omega^2}{v^2}  > 0 \nonumber \\
\lambda_3  &=&\frac{m_h^2}{v^2} - \frac{1}{s_{\beta}c_{\beta}}\, \frac{\Omega^2}{v^2} > - \sqrt{\lambda_1 \lambda_2}  \\
\lambda_3 + \lambda_4 - \left|\lambda_5  \right| &=& \frac{m_h^2}{v^2}  + \frac{\Delta^2}{v^2} - \left| \frac{1}{s_{\beta}c_{\beta}}\, 
\frac{\Omega^2}{v^2} - \frac{\Delta^2}{v^2}\right| > -\sqrt{\lambda_1 \lambda_2} \nonumber
\eea}

The first three inequalities in (\ref{Stability2}) are trivially satisfied for $\Omega^2 \leq 0$. For $\left|\Delta^2\right| \gg v^2$ the last one however requires 
$\Omega^2 \sim s_{\beta} c_{\beta} \Delta^2$, and this affects the unitarity bounds which depend on $\lambda_1 + \lambda_2$, {\it e.g.} 
{\small 
\bea
\label{Unitarity3}
\left|\lambda_1 + \lambda_2 + \sqrt{(\lambda_1-\lambda_2)^2 + 4 \lambda_4^2} \right| & \sim & 2\, \frac{\left|\Delta^2\right|\,t^2_{\beta}}{v^2} \quad (t_{\beta} \gg 1)\nonumber \\
& \sim & 2\, \frac{\left|\Delta^2\right|}{v^2\,t^2_{\beta}} \quad (t_{\beta} \ll 1)
\eea
}
such that only values $t_{\beta} \sim 1$ are allowed if $\left|\Delta^2\right| \gg v^2$.  
 
\vspace{-3mm} 

\subsection{Comments on $H^{\pm}$ Searches at LHC}
\label{Hpsection}

\vspace{-3mm}

Before we comment on the limits from direct searches of $H^{\pm}$, let us emphasize that there are two other important sources of constraints on the mass of 
$H^{\pm}$ in this case: {\it (i)} Flavour Physics yields important bounds on $m_{H^{\pm}}$, the most stringent one coming from the $H^{\pm}$ 
contribution to the flavour violating decay $b \to s \gamma$. For Type II 2HDM, this leads to a lower bound $m_{H^{\pm}} > 480$ GeV at 
95\% C.L.~\cite{Misiak:2015xwa}, while for Type I the bound is milder and depends on $t_{\beta}$~\cite{Hermann:2012fc}. 
{\it (ii)} EWPO strongly prefer $m_{H^{\pm}} \sim m_{A_0}$ or $m_{H^{\pm}} \sim m_{H_0}$ (this last condition is mildly modified away from the 
alignment limit $c_{\beta- \alpha} = 0$), as a splitting between the charged and neutral components of the doublet breaks custodial symmetry. While some degree of splitting
is allowed by EWPO, it cannot be sizable (see {\it e.g.}~the analysis of~\cite{Bernon:2015qea}). 
In the present work we have chosen for simplicity to 
make $H^{\pm}$ degenerate with the heavier of the two neutral scalars $H_0$, $A_0$, as neglecting small mass splittings between the charged and neutral scalars
does not have an appreciable impact on the theoretical constraints on the model, nor on the phenomenological analysis, and satisfies EWPO.
Regarding the bounds from Flavour Physics, while particularly for Type II they motivate our choice of pairing $H^{\pm}$ with the heavier state among 
$H_0$, $A_0$, we do not consider them as limits {\it stricto senso}, meaning that for Type II we still discuss scenarios in which both $m_{A_0}$ and $m_{H_0}$ 
are below $480$~GeV. We also stress that, since for Type I the $b \to s \gamma$ bound is not as severe, it could be possible for $H^{\pm}$ to pair 
with the lighter state among $H_0$, $A_0$. For a hierarchical 2HDM scenario, this would also 
open either the decay $A_0 \to W^{\pm} H^{\pm}$ or $H_0 \to W^{\pm} H^{\pm}$, and would make the LHC limits from searches of 
$A_0$, $H_0$ into SM states even weaker, opening at the same time further opportunities for direct searches of these new states (see {\it e.g.}~\cite{Coleppa:2014cca,Li:2015lra}).

\begin{widetext}
\onecolumngrid
\begin{figure}[h!]
\begin{center}
\vspace{-4mm}
\includegraphics[width=0.9\textwidth]{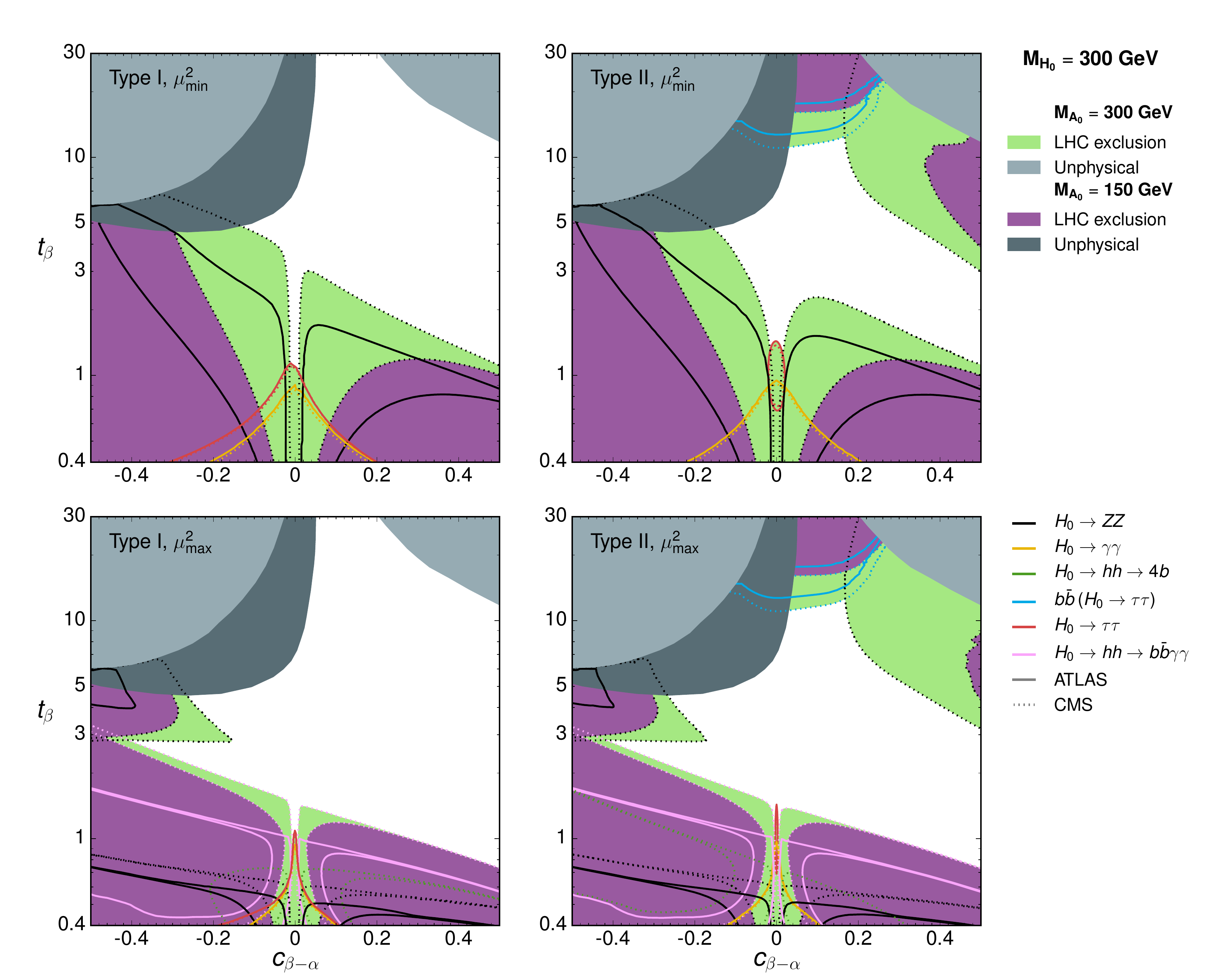}
\caption{\small 
Current $95 \,\%$ C.L. excluded region by ATLAS (solid lines) and CMS (dashed lines) for $m_{H_0} = 300$ GeV 
and respectively for 2HDM Type I/II (\textsl{Left}/\textsl{Right}) in the case $\mu^2 = \mu^2_{\mathrm{min}}$/$\mu^2 = \mu^2_{\mathrm{max}}$ (Top/Bottom). 
The limits come from searches of $g g, b \bar{b} \to H_0 \to WW,\,ZZ$ (black lines), $g g \to A_0 \to \tau \tau$ 
(red lines), $b \bar{b} \to A_0 \to \tau \tau$ (light-blue lines), $g g \to A_0 \to \gamma \gamma$ (yellow lines), 
$g g, b \bar{b} \to H_0 \to hh \to b \bar{b} b \bar{b}$ (dark-green lines) and $g g, b \bar{b} \to H_0 \to hh \to b \bar{b} \gamma\gamma$ (pink lines). 
The two scenarios considered are $m_{H_0} = m_{A_0}$ (Degenerate: green exclusion region) and $m_{H_0} - m_{A_0} = 150$ GeV (Hierarchical: purple exclusion region). 
The light/dark grey areas correspond to the theoretically excluded regions for the degenerate/hierarchical 2HDM scenarios.}
\label{fig:3}
\end{center}
\end{figure}

\end{widetext}

\begin{widetext}
\onecolumngrid

\begin{figure}[h!]
\begin{center}
\includegraphics[width=0.9\textwidth]{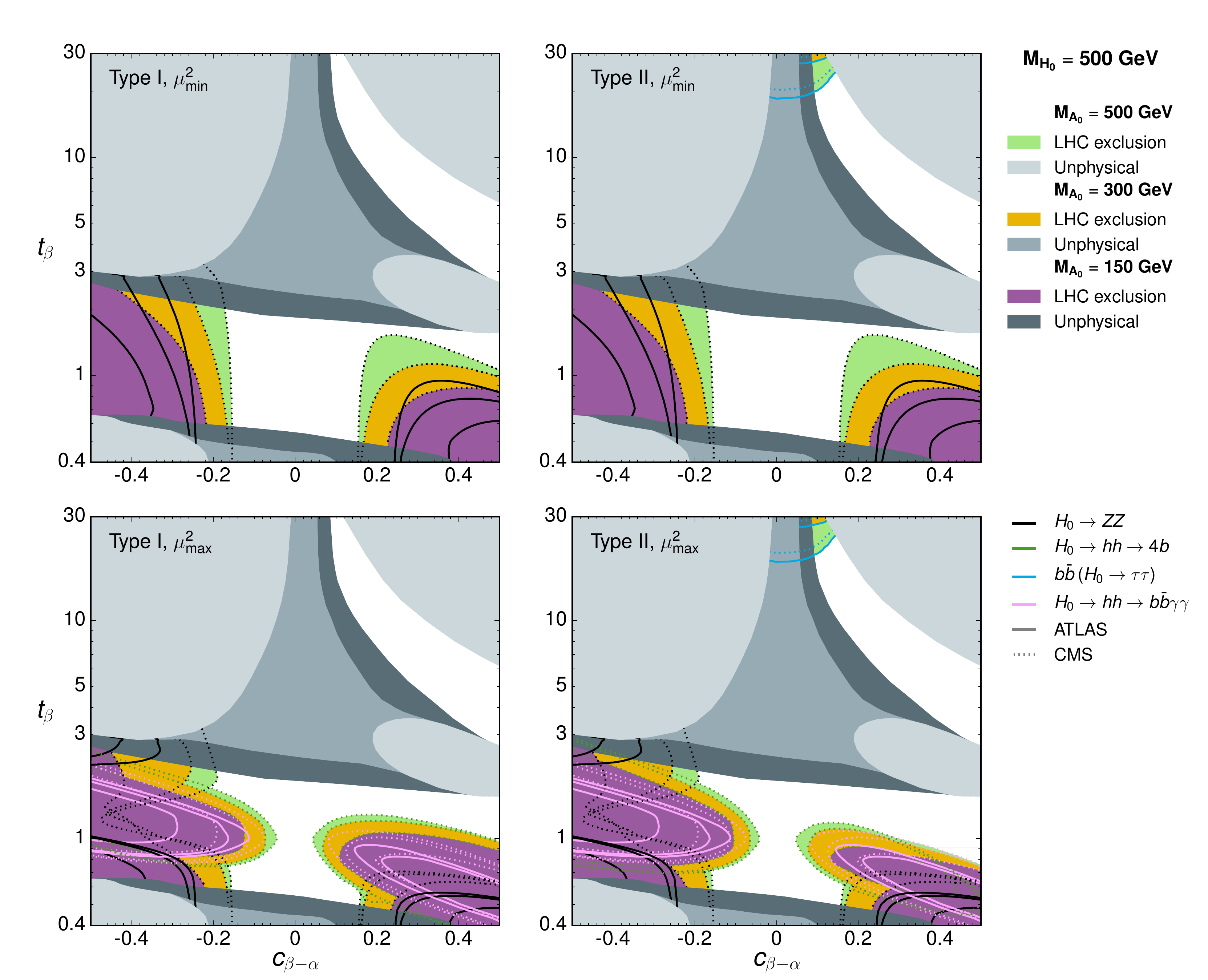}
\caption{\small 
Current $95 \,\%$ C.L. excluded region by ATLAS (solid lines) and CMS (dashed lines) for $m_{H_0} = 500$ GeV 
and respectively for 2HDM Type I/II (\textsl{Left}/\textsl{Right}) in the case $\mu^2 = \mu^2_{\mathrm{min}}$/$\mu^2 = \mu^2_{\mathrm{max}}$ (Top/Bottom). 
The limits come from searches of $g g, b \bar{b} \to H_0 \to WW,\,ZZ$ (black lines), $b \bar{b} \to A_0 \to \tau \tau$ (light-blue lines), 
$g g, b \bar{b} \to H_0 \to hh \to b \bar{b} b \bar{b}$ (dark-green lines) and $g g, b \bar{b} \to H_0 \to hh \to b \bar{b} \gamma\gamma$ (pink lines). 
The two scenarios considered are $m_{H_0} = m_{A_0}$ (Degenerate: green exclusion region) and $m_{H_0} - m_{A_0} = 150$ GeV (Hierarchical: purple exclusion region). 
The light/dark grey areas correspond to the theoretically excluded regions for the degenerate/hierarchical 2HDM scenarios.}
\label{fig:4}
\vspace{-4mm}
\end{center}
\end{figure}

\end{widetext}
 
We now briefly discuss the current bounds from searches of $H^{\pm}$ by ATLAS and CMS. For a light $H^{\pm}$, $m_{H^{\pm}} < m_t = 173$ GeV, ATLAS 
searches for $t \to H^{\pm}b$ in top quark pair production with the full dataset of Run 1 \cite{TheATLAScollaboration:2013wia,Aad:2014kga} set a 95\% C.L. bound on the 
branching fraction $\mathrm{BR}(t \to H^{\pm}b) \times \mathrm{BR}(H^{\pm} \to \nu\tau) < [0.0023,0.013]$  
in the mass range $m_{H^{\pm}}\in [80\,\mathrm{GeV}, 160\,\mathrm{GeV}]$.
For $m_{H^{\pm}} > m_t$, ATLAS searches for $H^{\pm}$ produced in association with a top quark \cite{Aad:2014kga} yield the bound 
$\sigma(pp \to tH^{\pm} + X) \times \mathrm{BR}(H^{\pm} \to \nu\tau) < [0.76\,\mathrm{pb},4.5\,\mathrm{fb}]$
in the range $m_{H^{\pm}}\in [180\,\mathrm{GeV}, 1000\,\mathrm{GeV}]$.
We however note that these bounds do not result generically in meaningful constraints, since 
$\mathrm{BR}(H^{\pm} \to \nu\tau) \ll 1$ when the decay $H^{\pm} \to tb$ is open. Moreover, 
in the hierarchical scenario $\mathrm{BR}(H^{\pm} \to \nu\tau)$ may be further suppressed by the presence of either 
$H^{\pm} \to W^{\pm}A_0$ or $H^{\pm} \to W^{\pm}H_0$ decays.



\subsection{Filling the Gaps: $A_0 \to Z H_0$/$H_0 \to Z A_0$ Searches}
\label{sec:bsm3}

\vspace{-3mm}

Our previous analysis highlights that, while direct searches for heavy neutral Higgs bosons at the LHC may provide a wide coverage across the 2HDM parameter space, and 
complementary to measurements of signal strengths, bounds from searches assuming direct decays of the neutral scalars into SM particles
become much weaker in a hierarchical 2HDM scenario, and new searches are needed to fill in the gaps.
It is also clear that the new searches capable of probing a hierarchical 2HDM are precisely those which exploit the sizable mass splittings among the neutral 
scalars, namely\footnote{Other decay modes could also be promising, like $A_0/H_0 \to W^{\pm} H^{\pm}$ or $H^{\pm} \to W^{\pm} A_0/H_0$, 
depending on $m_{H^{\pm}}$~\cite{Coleppa:2014cca,Li:2015lra})} $A_0 \to Z H_0$ or $H_0 \to Z A_0$. 
In the former case, the relevant final state to search for would depend on the dominant decay mode of $H_0$~\cite{Dorsch:2014qja}. For $c_{\beta - \alpha} \sim 0$, $H_0 \to b \bar{b}$ 
(eventually, $H_0 \to t \bar{t}$ if $m_{H_0} > 340$ GeV) would dominate, yielding $A_0 \to Z H_0 \to \ell \ell \,\,b \bar{b}$ as most sensitive final state. 
For a sizable departure from alignment, the dominant decay mode would be $H_0 \to W^{+} W^{-}$, yielding 
$A_0 \to Z H_0 \to \ell \ell \,\,W^{+} W^{-}$ ($W^{+} W^{-} \to \ell \nu\,\, \ell \nu $) as the most sensitive final 
state\footnote{Other competitive final states are $H_0 \to W^{+} W^{-}$ ($W^{+} W^{-} \to \ell \nu\,\, j j$) and 
$H_0 \to Z Z$ yielding $A_0 \to Z H_0 \to \ell \ell \ell'\ell' j j$~\cite{Coleppa:2014hxa}.}.

\vspace{1mm}

For $H_0 \to Z A_0$, the most sensitive final state is generically $\ell \ell \,\,b \bar{b}$. 
In~\cite{CMS:2015mba}, the CMS Collaboration has performed the first analysis of such signatures, with an integrated luminosity of 
$\mathcal{L} = 19.8\,\mathrm{fb}^{-1}$ at 8 TeV, in the $\ell \ell \,\,b \bar{b}$ final state relevant
for both $A_0 \to Z H_0$ and $H_0 \to Z A_0$. We discuss here the limits on the 2HDM parameter space that may be derived from that search.
We stress that from the point of view of the CMS analysis, the limits on the production cross sections for $A_0 \to Z H_0$ and $H_0 \to Z A_0$
are identical for the same kinematical mass point. However, the translation between these limits and the constraints on the 2HDM parameter space
is quite different in the two cases. 

\begin{widetext}
\onecolumngrid

\begin{figure}[h!]
\begin{center}
\vspace{-2mm}
\includegraphics[width=0.99\textwidth]{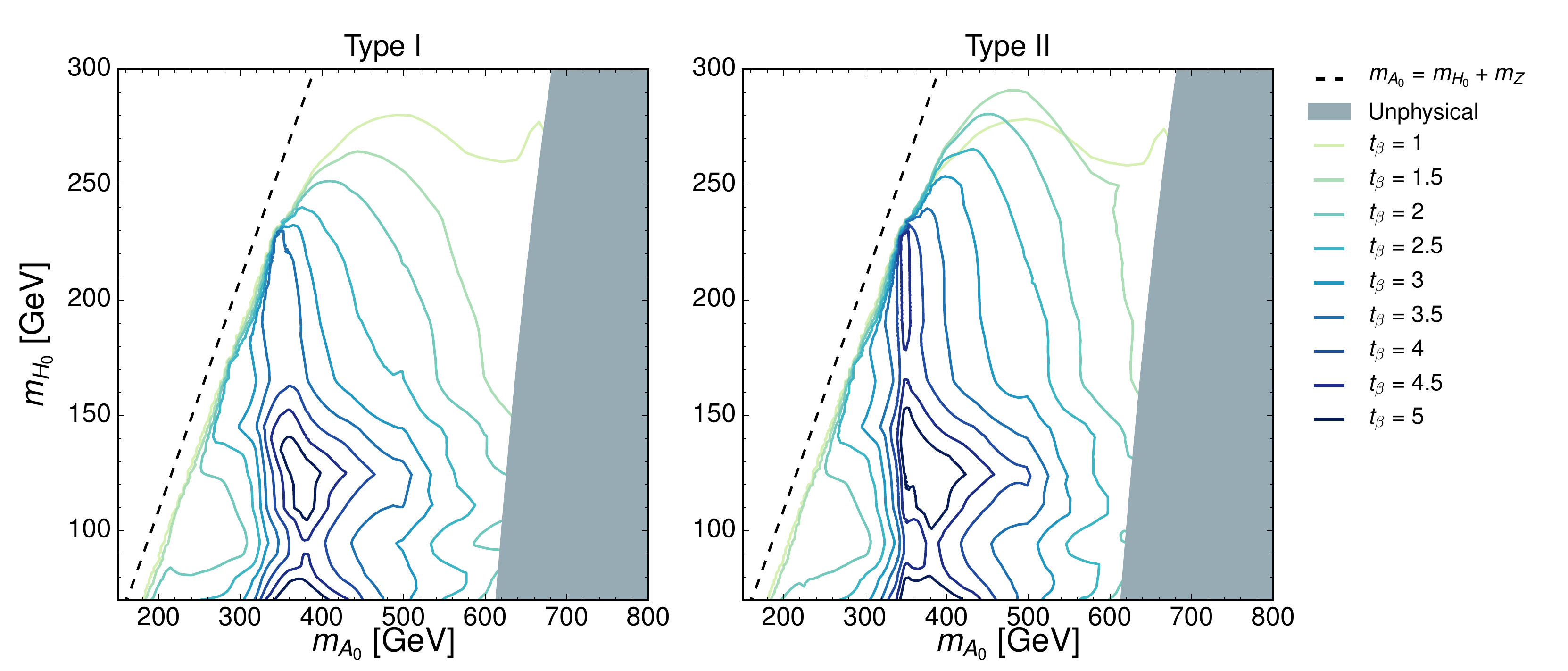}
\caption{\small Bounds on $t_{\beta}$ in the ($m_{A_0},\,m_{H_0}$) plane for $c_{\beta-\alpha} = 0$, from the search for $A_0 \to Z H_0 \to \ell \ell \,\,b \bar{b}$ 
performed in \cite{CMS:2015mba}, For 2HDM of Type I (\textsl{Left}) and Type II (\textsl{Right}). The grey regions 
are theoretically excluded.}
\label{fig:7}
\vspace{-3mm}
\end{center}
\end{figure}

\end{widetext}

Concentrating first on $A_0 \to Z H_0$, we show in Figure \ref{fig:7} the bounds on 
$t_{\beta}$ in the ($m_{A_0},\,m_{H_0}$) plane for Type I (\textsl{Left}) and Type II (\textsl{Right}), assuming $c_{\beta-\alpha} = 0$.
The search constrains up to $t_{\beta} \sim 5$ around $m_{A_0} = 380$ GeV, and additionally yields the limit $t_{\beta} \gtrsim 2$ 
for $m_{H_0} < 80$ GeV and $m_{A_0} < 600$ GeV. 
However, we expect a weakening of these limits once there is departure from the alignment limit, and we emphasize that 
searches for the $H_0 \to W^{+} W^{-}$ decay mode (and $Z Z$) are very much needed in this region (note that for $t^{-1}_{\beta}s_{\beta-\alpha} - c_{\beta-\alpha} \sim 0$, 
direct searches for $H_0 \to W^{+} W^{-}$ assuming gluon fusion production will not be sensitive to $H_0$).

\vspace{2mm}

In order to illustrate the complementarity between the above limits from $A_0 \to Z H_0$ searches and those from the most 
sensitive ATLAS/CMS searches for $A_0$, $H_0$ decaying directly into SM states analyzed in Sections \ref{A0section} and \ref{H0section}, as well as their interplay with  
measurements of Higgs signal strengths from Section \ref{sec:HSs} (we take here the limits obtained with {\sc HiggsSignals}), 
we present a summary of the various bounds on the ($c_{\beta-\alpha}$, $t_{\beta}$) plane 
in Figures \ref{fig:8} and \ref{fig:9} for Type I/II (\textsl{Left}/\textsl{Right}): Figure \ref{fig:8} (Top)
shows the combined limits for $m_{A_{0}} = m_{H_0} = 150$ GeV (only the degenerate scenario is considered in this case). 
Focusing then on $m_{A_{0}} = 300$ GeV, Figure \ref{fig:8} (Bottom) highlights the fact that 
close to $c_{\beta-\alpha} = 0$ the CMS search for $A_0 \to Z H_0$ ($H_0 \to \bar{b} b$) in the hierarchical 2HDM scenario 
yields a superior sensitivity to the one obtained in the degenerate 2HDM scenario
via the union of limits from $A_0$ and $H_0$ searches. It is also interesting to note that 
while in the degenerate scenario the combination of $A_0$ and $H_0$ searches 
exclude the Type II {\it wrong-sign} region allowed by Higgs signal strength measurements, 
in the hierarchical scenario the {\it wrong-sign} region is allowed by direct searches. 

\vspace{2mm}


We note that for $m_{A_{0}} = 150$ GeV and $m_{A_{0}} = 300$ GeV the choice between $\mu^2 = \mu^2_{\mathrm{min}}$ and $\mu^2 = \mu^2_{\mathrm{max}}$
does not impact the limits shown in Figure \ref{fig:8} since the di-Higgs searches are not the most constraining in this plane. For $m_{A_{0}} = 500$ GeV the 
situation is different, as shown in Figure \ref{fig:9}. Here, the degenerate case is the least constrained, cutting into the edges of the Type I light Higgs 
limits and not significantly affecting the Type II exclusions near alignment. In the $\mu_{\text{max}}$ scenario, the di-Higgs searches improve the limits 
towards alignment around $t_\beta\sim 1$. As one decreases the $H_0$ mass the picture changes considerably, with the direct $H_0$ searches proving particularly 
effective for $m_{H_0}=300$ GeV, even near alignment. For $m_{H_0}=150$ GeV, the $A_0\to Z\,H_0$ becomes sensitive and provides excellent coverage up to 
$t_\beta\sim 3$, generally improving on the direct searches. We note, again, that the {\it wrong-sign} scenario in Type II is excluded in the degenerate 
and $m_{H_0}=300$ GeV cases, while it is mostly allowed in the lightest $H_0$ case. 

\begin{widetext}
\onecolumngrid

\begin{figure}[h!]
\begin{center}
\vspace{-2mm}
\includegraphics[width=0.92\textwidth]{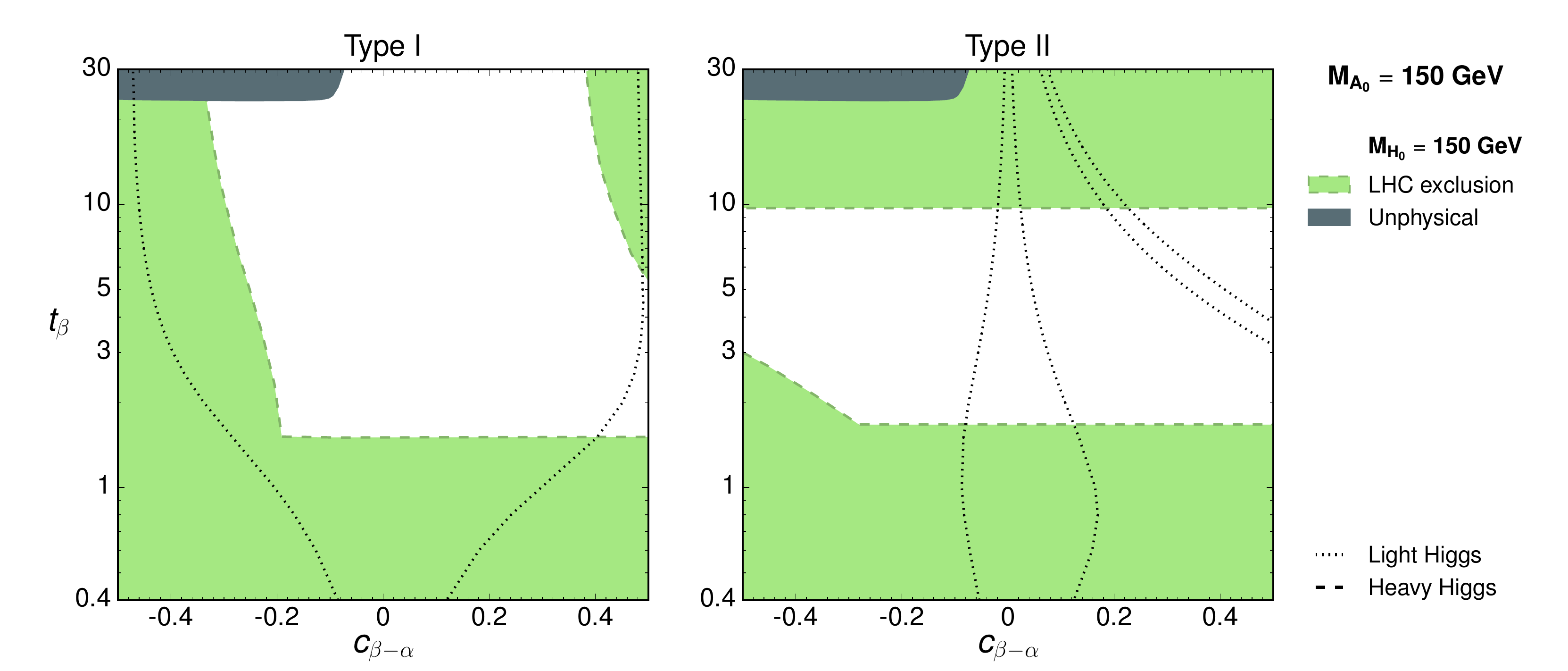}
\includegraphics[width=0.92\textwidth]{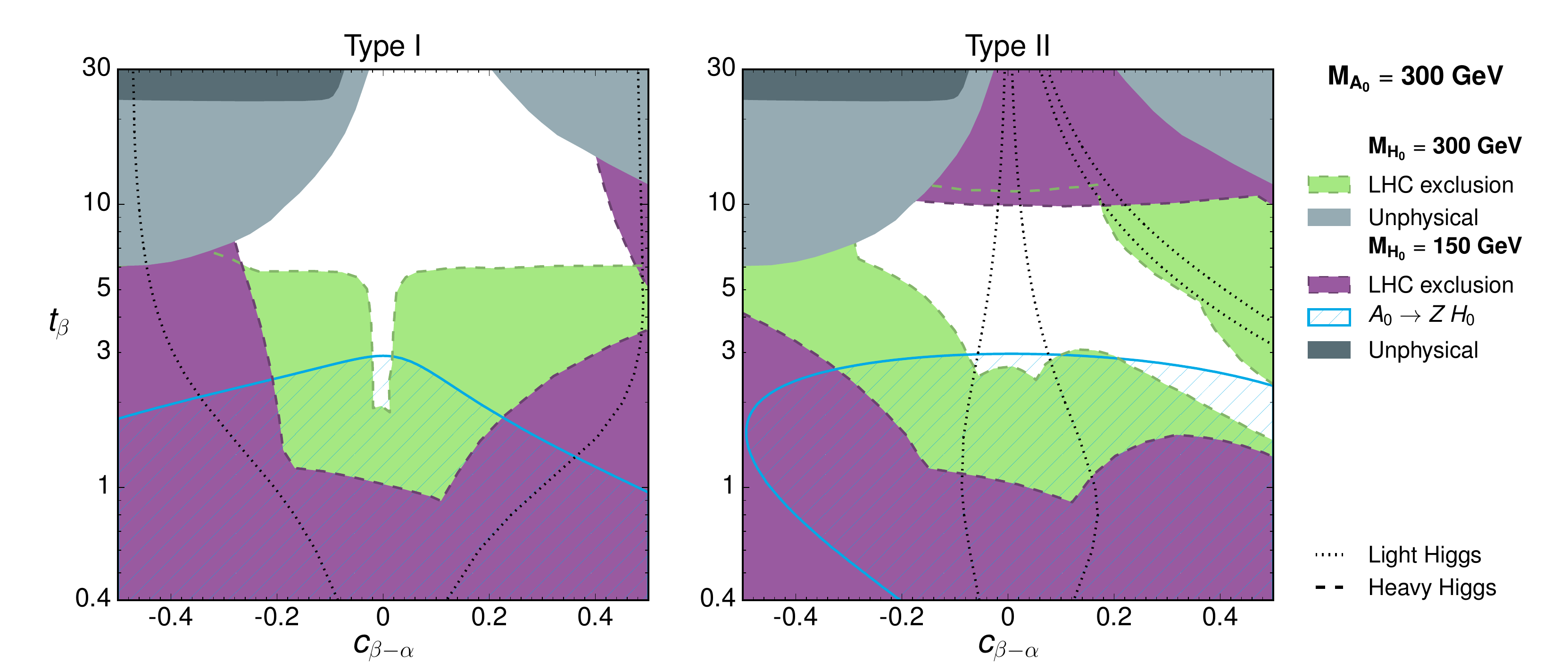}
\caption{\small 
Limits in the ($c_{\beta-\alpha}$, $t_{\beta}$) plane for $m_{A_0} = 150/300$~GeV (Top/Bottom) and Type I/II (\textsl{Left}/\textsl{Right}), from 
measurements of Higgs signal strengths obtained with {\sc HiggsSignals} (dotted black lines; see Section \ref{sec:HSs}) and from 
the most sensitive ATLAS/CMS searches for $A_0$ and $H_0$ decaying directly into SM states: 
green region corresponds to the exclusion in the degenerate scenario $m_{H_0} = m_{A_0}$; purple regions correspond to 
the exclusion in the hierarchical scenario (see Sections \ref{A0section}, \ref{H0section}).
The dashed blue region corresponds to the exclusion from the CMS $A_0 \to Z H_0 \to \ell\ell\, b \bar{b}$ search \cite{CMS:2015mba} in the hierarchical scenario. 
The grey regions are theoretically excluded.}
\label{fig:8}
\vspace{-2mm}
\end{center}
\end{figure}

\end{widetext}


\begin{widetext}
\onecolumngrid

\begin{figure}[h!]
\begin{center}
\vspace{-4mm}
\includegraphics[width=0.92\textwidth]{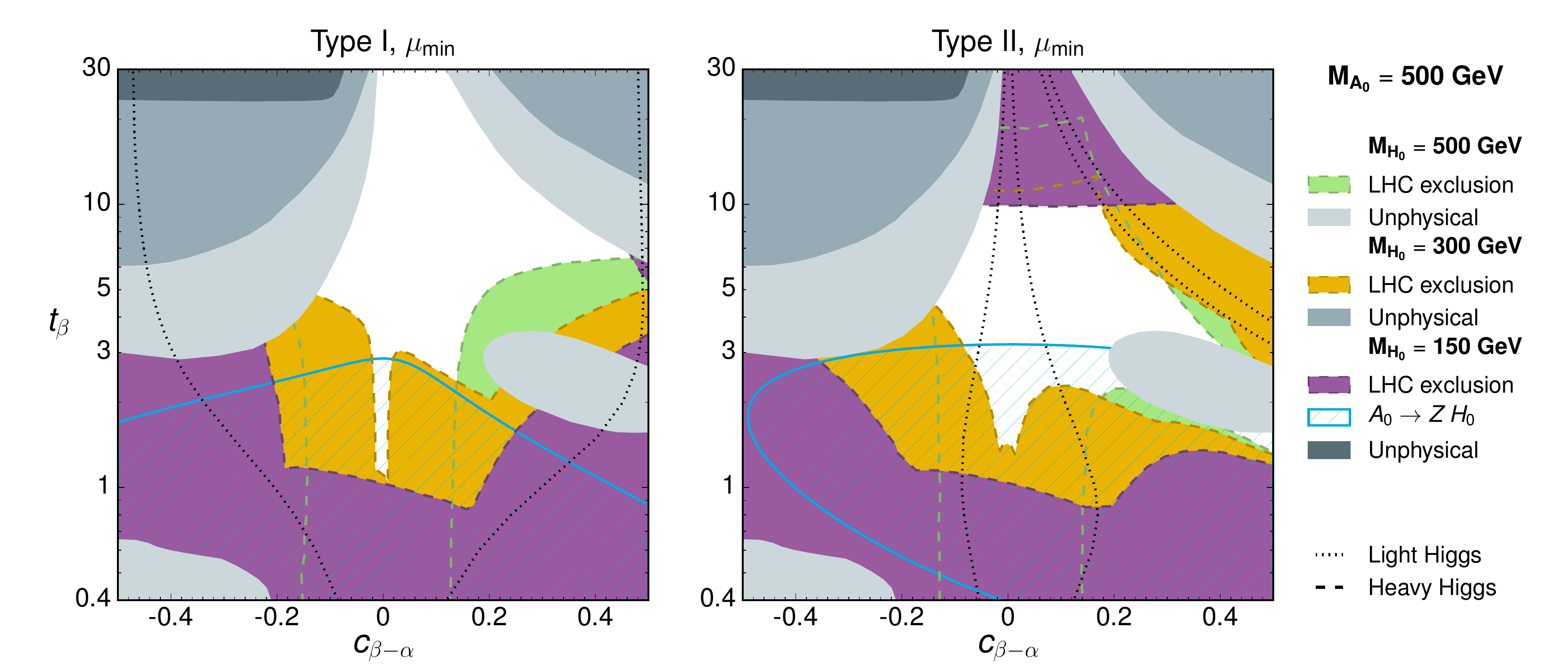}
\includegraphics[width=0.92\textwidth]{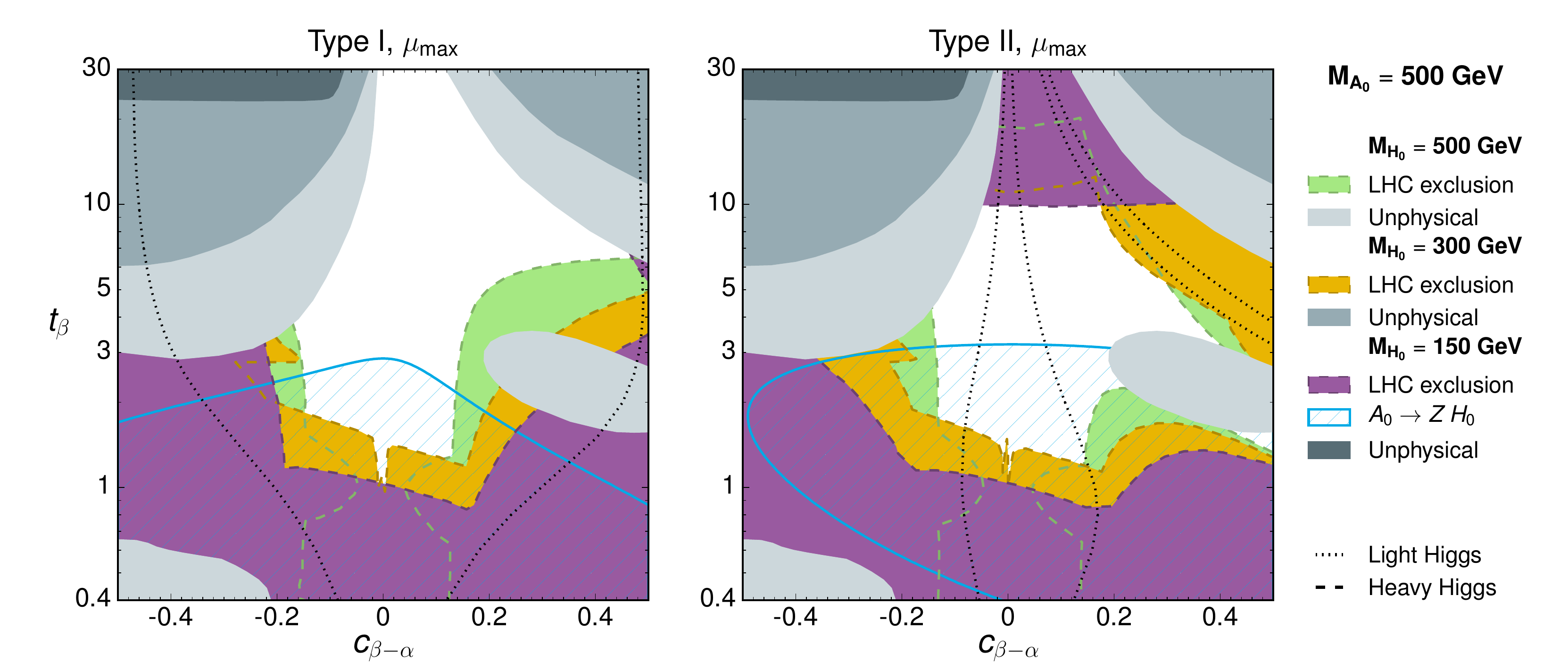}
\caption{\small 
Limits in the ($c_{\beta-\alpha}$, $t_{\beta}$) plane for $m_{A_0} = 500$~GeV with $\mu^2 = \mu^2_{\mathrm{min}}$/$\mu^2 = \mu^2_{\mathrm{max}}$ 
(Top/Bottom) and Type I/II (\textsl{Left}/\textsl{Right}), from measurements of Higgs signal strengths obtained with {\sc HiggsSignals} 
(dotted black lines; see Section \ref{sec:HSs}) and from the most sensitive ATLAS/CMS searches for $A_0$ and $H_0$ decaying directly into SM states: 
green region corresponds to the exclusion in the degenerate scenario $m_{H_0} = m_{A_0}$; purple/orange regions correspond to 
the exclusion in the hierarchical scenario(s) (see Sections \ref{A0section}, \ref{H0section}). 
The dashed blue region corresponds to the exclusion from the CMS $A_0 \to Z H_0 \to \ell\ell\, b \bar{b}$ search \cite{CMS:2015mba} in the hierarchical scenario
$m_{H_0} = 150$ GeV. The grey regions are theoretically excluded.}
\label{fig:9}
\vspace{-4mm}
\end{center}
\end{figure}

\end{widetext}

Turning to $H_0 \to Z A_0$, we show the limits on $t_{\beta}$ in the ($m_{A_0},\,m_{H_0}$) plane in Figure \ref{fig:10}, for
Type I (\textsl{Left}) and Type II (\textsl{Right}) in the alignment limit. A few comments are in order: First, the limits are expected to be weaker than for  
$A_0 \to Z H_0$, as the production cross section for $A_0$ is larger than that for $H_0$ for the same mass.~More importantly, when 
$m_{H_0} > 2\, m_{A_0}$, the decay $H_0 \to A_0 A_0$ becomes kinematically possible, which weakens the bounds from $H_0 \to Z A_0$ and also 
makes them dependent on $\mu^2$, since $\mathrm{BR}(H_0 \to A_0 A_0)$ does depend on this parameter. Figure~\ref{fig:10} (Top) shows the limits for 
$\mu^2 = \mu^2_{\mathrm{min}}$, while Figure~\ref{fig:10} (Bottom) shows the limits for $\mu^2 = \mu^2_{\mathrm{max}}$ which are identical 
for $m_{H_0} < 2\, m_{A_0}$ but much weaker for $m_{H_0} > 2\, m_{A_0}$ as expected. Note also that for $t_{\beta} = 1$ the limits are identical 
in both mass regions, since $\lambda_{H_0 A_0 A_0} = 0$ and the $\mu^2$ dependence therefore disappears. In the region $m_{H_0} > 2\, m_{A_0}$, the 
width of $H_0$ very quickly reaches $\Gamma_{H_0}/m_{H_0} > 0.15$, for which the bounds from the analysis~\cite{CMS:2015mba} are no longer robust (these regions
are marked as shaded in Figure \ref{fig:10}). This is in contrast with $A_0 \to Z H_0$ bounds, for which $\Gamma_{A_0}/m_{A_0} < 0.15$ throughout
the whole allowed parameter space. From the comparison of Figures \ref{fig:7} and \ref{fig:10} it is also apparent that 
for a sizable splitting $m_{H_0} - m_{A_0} > 0$, the 2HDM parameter space is much more 
theoretically constrained than for a splitting $m_{A_0} - m_{H_0} > 0$ of same magnitude, and the 
constraints become more stringent as $t_{\beta}$ increases (recall the discussion in Section \ref{H0section}), such that for $t_{\beta} > 2$
$m_{H_0} \lesssim 500$ GeV is required, as shown in Figure \ref{fig:10}.

\begin{widetext}
\onecolumngrid

\begin{figure}[h!]
\begin{center}
\vspace{-4mm}
\includegraphics[width=0.92\textwidth]{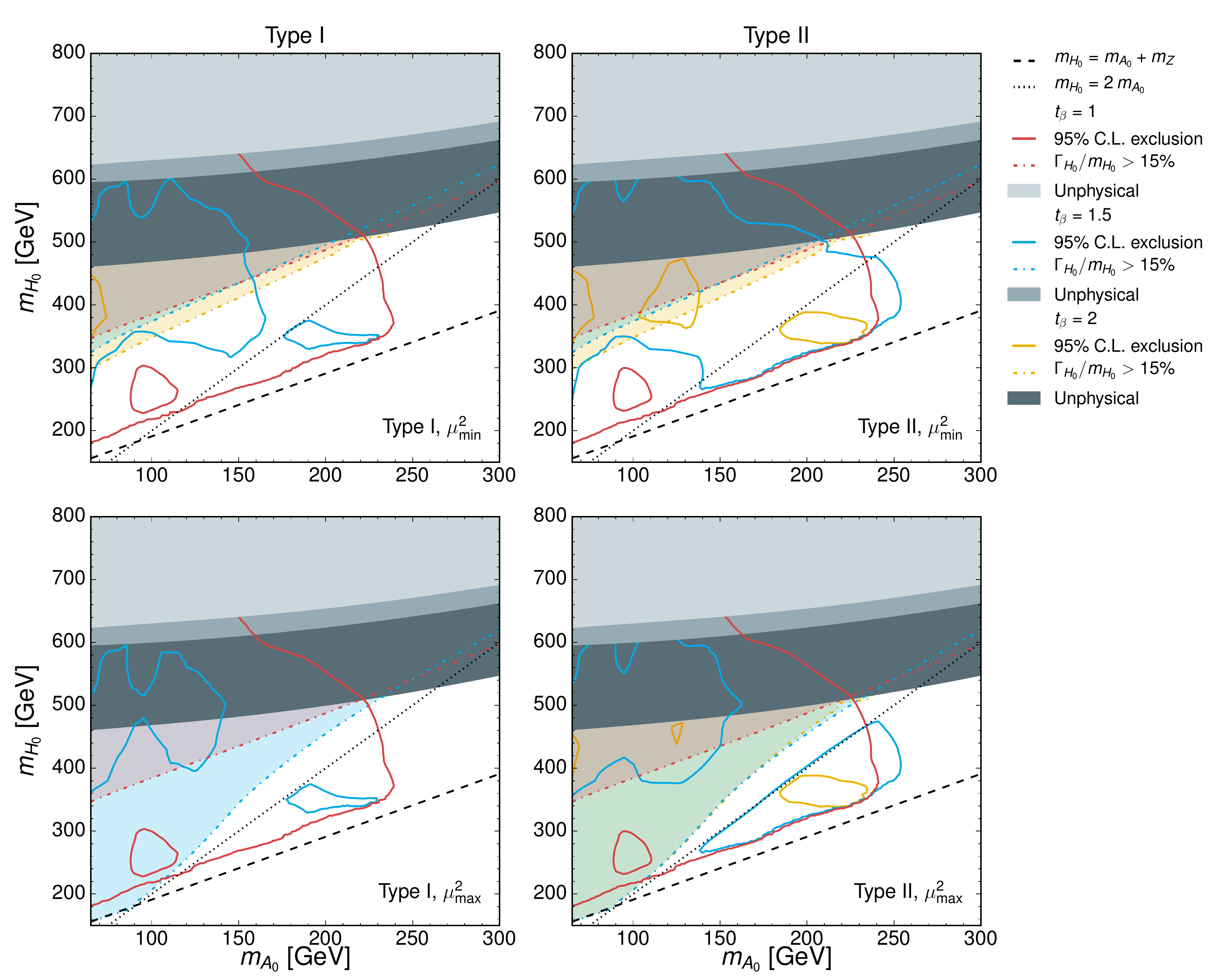}
\caption{\small Bounds on $t_{\beta}$ in the ($m_{A_0},\,m_{H_0}$) 
plane for $c_{\beta-\alpha} = 0$, from the search for $H_0 \to Z A_0 \to \ell \ell \,\,b \bar{b}$ performed in \cite{CMS:2015mba}, For 2HDM of Type 
I (\textsl{Left}) and Type II (\textsl{Right}), and for $\mu^2 = \mu^2_{\mathrm{min}}$/$\mu^2 = \mu^2_{\mathrm{max}}$ (Top/Bottom respectively). The grey regions 
are theoretically excluded, 
while the shaded regions correspond to $\Gamma_{H_0}/m_{H_0} > 0.15$. The dotted-black line corresponds to $m_{H_0} = 2\, m_{A_0}$, above which
the decay $H_0 \to A_0 A_0$ becomes kinematically possible.}
\label{fig:10}
\vspace{-4mm}
\end{center}
\end{figure}

\end{widetext}

\section{Hierarchical 2HDM and LHC Run II}
\label{sec:bsmRun2}

\vspace{-3mm}

LHC Run 2 at 13 TeV represents a great opportunity to dig further into the parameter space of hierarchical 2HDM scenarios, since the sensitivity of 
the searches described in the previous Section is limited mainly by small cross section values at the 8 TeV run of the LHC. 
While a detailed analysis of the LHC Run 2 prospects for the hierarchical scenario of the 2HDM is beyond the scope of this work, 
we present in this Section benchmark planes in ($m_{A_0},\,m_{H_0}$) for $A_0 \to Z H_0$ searches, 
classified according to the 2HDM Type (I/II) and the proximity to the alignment limit.
In Figure \ref{fig:11} (Top) we provide $\sigma(gg \to A_0 \to Z H_0)\times\mathrm{BR}(H_0 \to X)$ for Type I/II (\textsl{Left}/\textsl{Right}) and a reference value 
$t_{\beta} = 3$, with $X$ being the relevant decay mode of $H_0$ in each case: In alignment $c_{\beta-\alpha} = 0$ (Figure \ref{fig:11}, Top), $X$ is the main 
fermionic decay of $H_0$, namely $\bar{b} b$ for $m_{H_0} < 340$ GeV and $\bar{t} t$ for $m_{H_0} > 340$ GeV. 
Away from alignment $c_{\beta-\alpha} \gtrsim 0.2$ (Figure \ref{fig:11}, Bottom), $X = W^+ W^-$ and we choose $c_{\beta-\alpha} = 0.3$ for Type I, $c_{\beta-\alpha} = 0.5$ 
for Type II.
We show in each case the constraint from the LHC Run 1 $A_0 \to Z H_0$ ($H\to \bar{b} b$) CMS search \cite{CMS:2015mba}, noting that besides providing useful limits 
in alignment (recall Figure \ref{fig:7}), it can also constrain the ($m_{A_0},\,m_{H_0}$) plane away from alignment. 
This is most relevant in Type II, where $\kappa^{H_0}_d$ increases with $t_{\beta}$, and for $m_{H_0} \lesssim 180$ GeV 
as shown in Figure \ref{fig:11} (see also Figures \ref{fig:8} and  \ref{fig:9}). In contrast, for the benchmarks chosen away from alignment there are no limits from 
$g g \to H_0 \to W^+ W^-$ searches in the whole ($m_{A_0},\,m_{H_0}$) plane: for Type I this is due to $\kappa^{H_0}_u \ll 1$ ($H_0$ is approximately fermiophobic), 
while for Type II it is due to the $(\kappa^{H_0}_d)^2$ enhancement of the partial width $\Gamma(H_0 \to \bar{b} b)$ {\it vs} the $(\kappa^{H_0}_V)^2$ suppression
of the partial width $\Gamma(H_0 \to W^+ W^-)$.
The discussion above emphasizes the search $gg \to A_0 \to Z H_0$ ($H_0 \to W^+ W^-$) as potentially key to probe a hierarchical 2HDM scenario away from the alignment limit.

\begin{widetext}
\onecolumngrid

\begin{figure}[h!]
\begin{center}
\vspace{-1mm}
\includegraphics[width=0.97\textwidth]{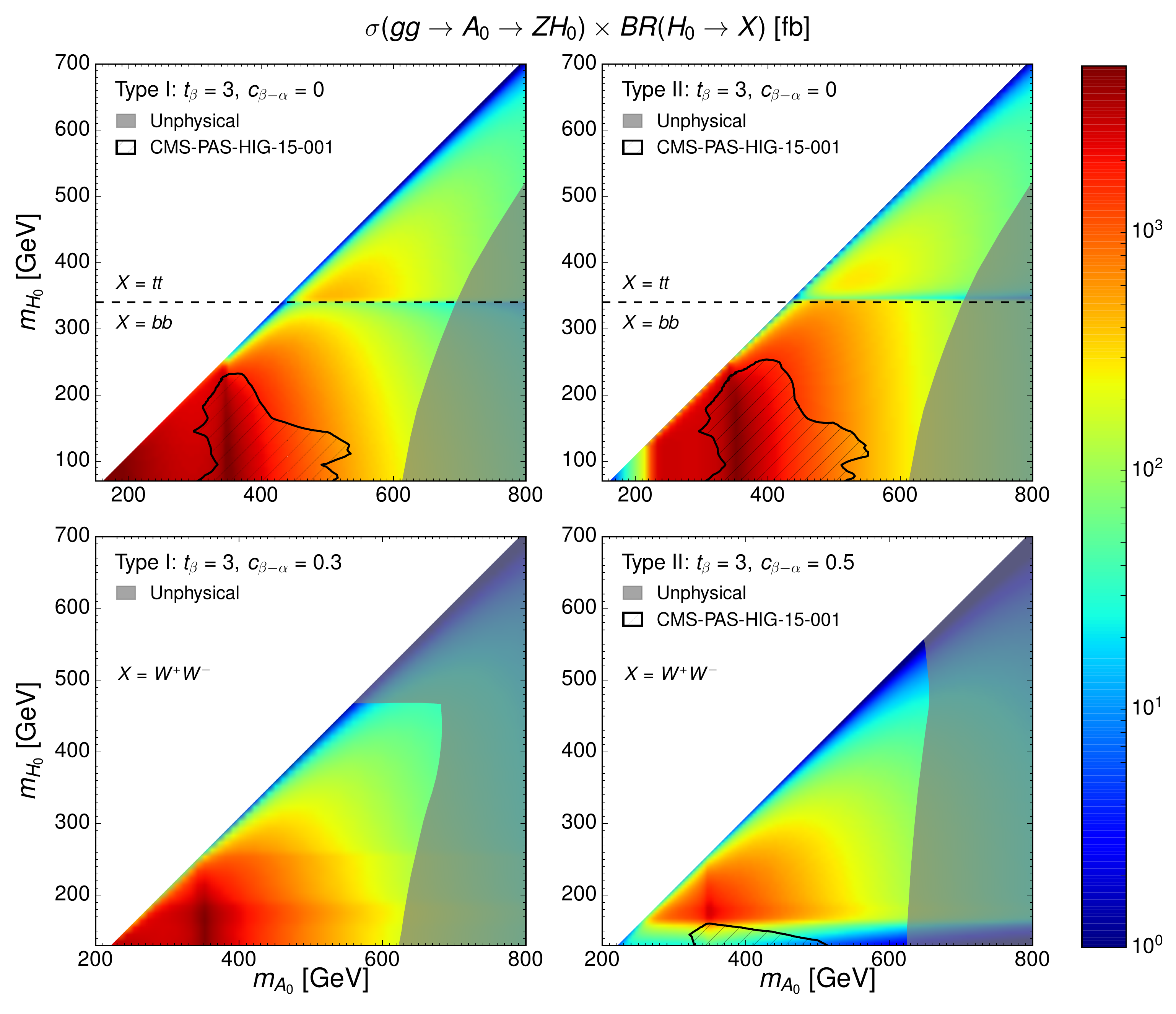} 
\caption{\small Cross section $\sigma(gg \to A_0 \to Z H_0) \times \mathrm{BR}(H_0 \to X)$ in the ($m_{A_0},\,m_{H_0}$) plane, for Type I (\textsl{Left}) and 
Type II (\textsl{Right}). Top: Alignment limit $c_{\beta-\alpha} = 0$, with $ X = \bar{b} b$ if $m_{H_0} < 340$ GeV and 
$ X = \bar{t} t$ if $m_{H_0} > 340$ GeV. The dashed black region corresponds to the exclusion from the LHC Run 1 $A_0 \to Z H_0$ CMS search \cite{CMS:2015mba}.
Bottom: Departure from alignment, with $ X = W^+ W^-$ and $c_{\beta-\alpha} = 0.3$ for Type I, $c_{\beta-\alpha} = 0.5$ 
for Type II. The value of the soft $\mathbb{Z}_2$-breaking parameter is fixed to $\mu^2=m_{H_0}^2\,s_\beta\,c_\beta$ ($\Omega^2=0$, see discussion in Section~\ref{A0section}). The grey regions are theoretically excluded.}
\label{fig:11}
\vspace{-1mm}
\end{center}
\end{figure}

\end{widetext}

Before concluding this section, a few comments are important: For Type II, the combination of Flavour bounds on $m_{H^{\pm}}$ and EWPO
would disfavour a 2HDM spectrum with both $m_{A_0}$ and $m_{H_0}$ significantly below $480$ GeV, as discussed in Section \ref{Hpsection}.
We choose not to show this in Figure \ref{fig:11}, as these indirect limits (particularly the Flavour bound) could be modified in the presence of new physics. 
Also, while we do not discuss here the prospects for searches of $H_0$ decaying into non-SM states, we emphasize that searches for $H_0 \to Z A_0$ and 
$H_0 \to A_0 A_0$ may be key to probe a hierarchical 2HDM scenario with $m_{H_0} > m_{A_0}$.


\section{Discussion and Outlook}
\label{sec:discuss}

\vspace{-3mm}

Uncovering the full structure of the SM scalar sector and its possible extensions will be a central task for the LHC in the coming years. 
The results will have important implications not only for our understanding of the mechanism of electroweak symmetry-breaking but also for the origin 
of visible matter and the nature of dark matter. Extensions of the SM scalar sector that address one or both of these open questions may yield distinctive 
signatures at the LHC via modifications of the SM Higgs boson properties and/or the observation of new states.

In this work we have investigated the constraints on the parameter space of CP-conserving two-Higgs-doublet models of Types I/II in light of the ATLAS/CMS results from LHC Run 1. 
A key difference from the many similar analyses already existing in the literature is that the 
latter generally assume a nearly degenerate 2HDM spectrum for the new scalar states, which can then only decay into SM particles. 
While the properties of the observed 125 GeV Higgs are not affected by the mass spectrum of the new scalars (as discussed in Section \ref{sec:HSs}), 
a large mass splitting between two or more of the new scalar states, \emph{e.g.} $m_{A_0}-m_{H_0}\gtrsim m_Z$, 
causes new decay channels of the heavier scalars to open and become dominant. For such a \emph{hierarchical} 2HDM,
we show that the constraints usually obtained in the literature are \emph{significantly} weakened.
On the other hand, the new decay channels constitute novel ways of searching for these scalar states, \emph{e.g.} $A_0\to ZH_0$
and we show how they can be used to fill in the gaps left by previous analyses.
We also highlight the importance of the $\mu^2$ parameter, through its impact on the phenomenology of the heavier CP-even 
scalar $H_0$ and its sensitivity to unitarity and stability constraints.



\vspace{2mm}


Finally, we believe this analysis will strongly contribute to provide a global picture of the present status of 2HDMs in light of Run 1 LHC results, and of the prospects
and relevant searches needed for exploring the still unconstrained regions of their parameter space.

\begin{center}
\textbf{Acknowledgements} 
\end{center}

We thank Jeremy Bernon for useful discussions. The work of S.H and K.M. is supported by the Science Technology and Facilities Council (STFC) under grant number
ST/L000504/1. J.M.N. is supported by the People Programme (Marie curie Actions) of the European Union Seventh Framework Programme (FP7/2007-2013) under REA 
grant agreement PIEF-GA-2013-625809. G.C.D. is supported by the German Science Foundation (DFG) under the Collaborative Research Center (SFB) 676 Particles, Strings and
the Early Universe.


\begin{thebibliography}{99}

\bibitem{Aad:2015gba}
  G.~Aad {\it et al.} [ATLAS Collaboration],
  arXiv:1507.04548 [hep-ex].

\bibitem{Khachatryan:2014jba}
  V.~Khachatryan {\it et al.} [CMS Collaboration],
  Eur.\ Phys.\ J.\ C {\bf 75} (2015) 5,  212
  doi:10.1140/epjc/s10052-015-3351-7
  [arXiv:1412.8662 [hep-ex]].

\bibitem{Branco:2011iw} 
  G.~C.~Branco, P.~M.~Ferreira, L.~Lavoura, M.~N.~Rebelo, M.~Sher and J.~P.~Silva,
  Phys.\ Rept.\  {\bf 516}, 1 (2012)
  [arXiv:1106.0034 [hep-ph]]. 

\bibitem{Cline:1995dg} 
  J.~M.~Cline, K.~Kainulainen and A.~P.~Vischer,
  Phys.\ Rev.\ D {\bf 54}, 2451 (1996)
  [hep-ph/9506284].
  
\bibitem{Cline:1996mga} 
  J.~M.~Cline and P.~A.~Lemieux,
  Phys.\ Rev.\ D {\bf 55}, 3873 (1997)
  [hep-ph/9609240].
  
\bibitem{Fromme:2006cm} 
  L.~Fromme, S.~J.~Huber and M.~Seniuch,
  JHEP {\bf 0611}, 038 (2006)
  [hep-ph/0605242].  
  
\bibitem{Dorsch:2013wja}
G.~C.~Dorsch, S.~J.~Huber and J.~M.~No,
JHEP {\bf 1310} (2013) 029
[arXiv:1305.6610 [hep-ph]].

\bibitem{Dorsch:2014qja}
  G.~C.~Dorsch, S.~J.~Huber, K.~Mimasu and J.~M.~No,
  Phys.\ Rev.\ Lett.\  {\bf 113} (2014) 21,  211802
  [arXiv:1405.5537 [hep-ph]].
  
 
  \bibitem{Aad:2015pla} 
  G.~Aad {\it et al.} [ATLAS Collaboration],
  arXiv:1509.00672 [hep-ex].
   
  
\bibitem{Celis:2013rcs} 
  A.~Celis, V.~Ilisie and A.~Pich,
  JHEP {\bf 1307}, 053 (2013)
  [arXiv:1302.4022 [hep-ph]].

\bibitem{Krawczyk:2013gia} 
  M.~Krawczyk, D.~Sokolowska and B.~Swiezewska,
  J.\ Phys.\ Conf.\ Ser.\  {\bf 447}, 012050 (2013)
  [arXiv:1303.7102 [hep-ph]].

\bibitem{Grinstein:2013npa} 
  B.~Grinstein and P.~Uttayarat,
  JHEP {\bf 1306}, 094 (2013)
  [Erratum-ibid.\  {\bf 1309}, 110 (2013)]
  [arXiv:1304.0028 [hep-ph]].
   
\bibitem{Chen:2013rba} 
  C.~-Y.~Chen, S.~Dawson and M.~Sher,
  Phys.\ Rev.\ D {\bf 88}, 015018 (2013)
  [arXiv:1305.1624 [hep-ph]].  

\bibitem{Craig:2013hca}
  N.~Craig, J.~Galloway and S.~Thomas,
  arXiv:1305.2424 [hep-ph].

\bibitem{Eberhardt:2013uba} 
  O.~Eberhardt, U.~Nierste and M.~Wiebusch,
  JHEP {\bf 1307}, 118 (2013)
  [arXiv:1305.1649 [hep-ph]].   
  
  \bibitem{Dumont:2014wha}
  B.~Dumont, J.~F.~Gunion, Y.~Jiang and S.~Kraml,
  Phys.\ Rev.\ D {\bf 90} (2014) 035021
  [arXiv:1405.3584 [hep-ph]].
  
  \bibitem{Bernon:2014vta} 
  J.~Bernon, B.~Dumont and S.~Kraml,
  Phys.\ Rev.\ D {\bf 90}, 071301 (2014)
  [arXiv:1409.1588 [hep-ph]].
  
 \bibitem{Craig:2015jba} 
  N.~Craig, F.~D'Eramo, P.~Draper, S.~Thomas and H.~Zhang,
  JHEP {\bf 1506}, 137 (2015)
  [arXiv:1504.04630 [hep-ph]].
    
  \bibitem{Bernon:2015qea} 
  J.~Bernon, J.~F.~Gunion, H.~E.~Haber, Y.~Jiang and S.~Kraml,
  Phys.\ Rev.\ D {\bf 92}, no. 7, 075004 (2015)
  [arXiv:1507.00933 [hep-ph]].
  
  
\bibitem{CMS:2015mba} 
  CMS Collaboration [CMS Collaboration],
  CMS-PAS-HIG-15-001.  
   
  
  \bibitem{Bernon:2014nxa} 
  J.~Bernon, J.~F.~Gunion, Y.~Jiang and S.~Kraml,
  Phys.\ Rev.\ D {\bf 91}, no. 7, 075019 (2015)
  [arXiv:1412.3385 [hep-ph]]. 
 
 
 \bibitem{Bernon:2015wef} 
  J.~Bernon, J.~F.~Gunion, H.~E.~Haber, Y.~Jiang and S.~Kraml,
  arXiv:1511.03682 [hep-ph].  
 
 
  \bibitem{Glashow:1976nt} 
  S.~L.~Glashow and S.~Weinberg,
  Phys.\ Rev.\ D {\bf 15}, 1958 (1977). 
  
  \bibitem{Maniatis:2006fs} 
  M.~Maniatis, A.~von Manteuffel, O.~Nachtmann and F.~Nagel,
  Eur.\ Phys.\ J.\ C {\bf 48}, 805 (2006)
  [hep-ph/0605184].  
  
  
  \bibitem{Grinstein:2015rtl} 
  B.~Grinstein, C.~W.~Murphy and P.~Uttayarat,
  arXiv:1512.04567 [hep-ph].
  
  \bibitem{Akeroyd:2000wc} 
  A.~G.~Akeroyd, A.~Arhrib and E.~M.~Naimi,
  Phys.\ Lett.\ B {\bf 490}, 119 (2000)
  [hep-ph/0006035].  
  
  \bibitem{Grimus:2007if} 
  W.~Grimus, L.~Lavoura, O.~M.~Ogreid and P.~Osland,
  J.\ Phys.\ G {\bf 35}, 075001 (2008)
  [arXiv:0711.4022 [hep-ph]].
 
 \bibitem{Coleppa:2014hxa} 
  B.~Coleppa, F.~Kling and S.~Su,
  JHEP {\bf 1409}, 161 (2014)
  [arXiv:1404.1922 [hep-ph]].
  
  
  \bibitem{Coleppa:2014cca} 
  B.~Coleppa, F.~Kling and S.~Su,
  JHEP {\bf 1412}, 148 (2014)
  [arXiv:1408.4119 [hep-ph]].
  
  
 \bibitem{Li:2015lra}
  T.~Li and S.~Su,
  arXiv:1504.04381 [hep-ph].  
   
  \bibitem{Bernon:2015hsa} 
  J.~Bernon and B.~Dumont,
  Eur.\ Phys.\ J.\ C {\bf 75}, no. 9, 440 (2015)
  [arXiv:1502.04138 [hep-ph]]. 
   
   
   
  \bibitem{Bechtle:2013xfa} 
  P.~Bechtle, S.~Heinemeyer, O.~Stål, T.~Stefaniak and G.~Weiglein,
  Eur.\ Phys.\ J.\ C {\bf 74}, no. 2, 2711 (2014)
  [arXiv:1305.1933 [hep-ph]].
  
  \bibitem{Bechtle:2014ewa} 
  P.~Bechtle, S.~Heinemeyer, O.~Stål, T.~Stefaniak and G.~Weiglein,
  JHEP {\bf 1411}, 039 (2014)
  [arXiv:1403.1582 [hep-ph]].
  
  
    
\bibitem{ATLAS:2013wla}
  [ATLAS Collaboration],
  ATLAS-CONF-2013-030, ATLAS-COM-CONF-2013-028. 
  
  \bibitem{Aad:2013wqa} 
  G.~Aad {\it et al.} [ATLAS Collaboration],
  Phys.\ Lett.\ B {\bf 726}, 88 (2013)
  [Phys.\ Lett.\ B {\bf 734}, 406 (2014)]
  [arXiv:1307.1427 [hep-ex]].
  
  \bibitem{TheATLAScollaboration:2013hia} 
  The ATLAS collaboration [ATLAS Collaboration],
  ATLAS-CONF-2013-075.
    
  
  \bibitem{Chatrchyan:2013iaa}
  S.~Chatrchyan {\it et al.}  [CMS Collaboration],
  JHEP {\bf 1401} (2014) 096
  [arXiv:1312.1129 [hep-ex]]. 
  
  
  \bibitem{CMS:2013xda} 
  CMS Collaboration [CMS Collaboration],
  CMS-PAS-HIG-13-017.
  
  
 \bibitem{Aad:2014eva}
  G.~Aad {\it et al.}  [ATLAS Collaboration],
  arXiv:1408.5191 [hep-ex].

\bibitem{Chatrchyan:2013mxa}
  S.~Chatrchyan {\it et al.}  [CMS Collaboration],
  Phys.\ Rev.\ D {\bf 89} (2014) 092007
  [arXiv:1312.5353 [hep-ex]].  
  
  
 \bibitem{Aad:2014eha}
  G.~Aad {\it et al.}  [ATLAS Collaboration],
  arXiv:1408.7084 [hep-ex].
  
\bibitem{Khachatryan:2014ira}
  V.~Khachatryan {\it et al.}  [CMS Collaboration],
  Eur.\ Phys.\ J.\ C {\bf 74} (2014) 10,  3076
  [arXiv:1407.0558 [hep-ex]].    
  
  
  
  
  
\bibitem{TheATLAScollaboration:2013lia}
  The ATLAS collaboration,
  ATLAS-CONF-2013-079, ATLAS-COM-CONF-2013-080.  
  
\bibitem{Chatrchyan:2013zna}
  S.~Chatrchyan {\it et al.}  [CMS Collaboration],
  Phys.\ Rev.\ D {\bf 89} (2014) 1,  012003
  [arXiv:1310.3687 [hep-ex]].  
  
  
\bibitem{ATLAS_tau}
  The ATLAS collaboration,
  ATLAS-CONF-2013-108, ATLAS-COM-CONF-2013-095.  
  
\bibitem{Chatrchyan:2014nva}
  S.~Chatrchyan {\it et al.}  [CMS Collaboration],
  JHEP {\bf 1405} (2014) 104
  [arXiv:1401.5041 [hep-ex]].  
  
  
\bibitem{Ferreira:2014naa} 
  P.~M.~Ferreira, J.~F.~Gunion, H.~E.~Haber and R.~Santos,
  Phys.\ Rev.\ D {\bf 89}, no. 11, 115003 (2014)
  [arXiv:1403.4736 [hep-ph]].  
  
  
  
  
  
   \bibitem{Aad:2015wra}
  G.~Aad {\it et al.}  [ATLAS Collaboration],
  Phys.\ Lett.\ B {\bf 744} (2015) 163
  [arXiv:1502.04478 [hep-ex]].  
  
\bibitem{Khachatryan:2015lba} 
  V.~Khachatryan {\it et al.} [CMS Collaboration],
  Phys.\ Lett.\ B {\bf 748}, 221 (2015)
  [arXiv:1504.04710 [hep-ex]].     
  
  
  \bibitem{Aad:2014ioa}
  G.~Aad {\it et al.}  [ATLAS Collaboration],
  Phys.\ Rev.\ Lett.\  {\bf 113} (2014) 17,  171801
  [arXiv:1407.6583 [hep-ex]].  
  
\bibitem{CMS:2014onr}
  CMS Collaboration [CMS Collaboration],
  CMS-PAS-HIG-14-006.  
  
\bibitem{Aad:2014vgg}
  G.~Aad {\it et al.}  [ATLAS Collaboration],
  JHEP {\bf 1411} (2014) 056
  [arXiv:1409.6064 [hep-ex]].  
  
\bibitem{Khachatryan:2014wca}
  V.~Khachatryan {\it et al.}  [CMS Collaboration],
  JHEP {\bf 1410} (2014) 160
  [arXiv:1408.3316 [hep-ex]].  
  
\bibitem{CMS:2013xfa}
  [CMS Collaboration],
  arXiv:1307.7135.    
  
 
\bibitem{Harlander:2012pb} 
  R.~V.~Harlander, S.~Liebler and H.~Mantler,
  Computer Physics Communications {\bf 184}, 1605 (2013)
  [arXiv:1212.3249 [hep-ph]].   
  
\bibitem{Eriksson:2009ws} 
  D.~Eriksson, J.~Rathsman and O.~Stal,
  Comput.\ Phys.\ Commun.\  {\bf 181}, 189 (2010)
  [arXiv:0902.0851 [hep-ph]].   
    
  
\bibitem{ATLAS:2013nma}
  [ATLAS Collaboration],
  ATLAS-CONF-2013-013, ATLAS-COM-CONF-2013-018.  
    

  \bibitem{ATLAS:2014aga}
  G.~Aad {\it et al.}  [ATLAS Collaboration],
  arXiv:1412.2641 [hep-ex].  
  
  
\bibitem{Aad:2014yja}
  G.~Aad {\it et al.}  [ATLAS Collaboration],
  Phys.\ Rev.\ Lett.\  {\bf 114} (2015) 8,  081802
  [arXiv:1406.5053 [hep-ex]].  
    
  
 \bibitem{Khachatryan:2015cwa}
  V.~Khachatryan {\it et al.}  [CMS Collaboration],
  arXiv:1504.00936 [hep-ex].  
  
\bibitem{CMS:2014ipa}
  CMS Collaboration [CMS Collaboration],
  CMS-PAS-HIG-13-032.  
  
\bibitem{Khachatryan:2015yea}
  V.~Khachatryan {\it et al.}  [CMS Collaboration],
  arXiv:1503.04114 [hep-ex].  
   
  
  
\bibitem{Misiak:2015xwa}
  M.~Misiak {\it et al.},
  Phys.\ Rev.\ Lett.\  {\bf 114} (2015) 22,  221801
  [arXiv:1503.01789 [hep-ph]].
  

\bibitem{Hermann:2012fc}
  T.~Hermann, M.~Misiak and M.~Steinhauser,
  JHEP {\bf 1211} (2012) 036
  [arXiv:1208.2788 [hep-ph]].   
  
  
\bibitem{TheATLAScollaboration:2013wia} 
  The ATLAS collaboration [ATLAS Collaboration],
  ATLAS-CONF-2013-090.
  
  
\bibitem{Aad:2014kga} 
  G.~Aad {\it et al.} [ATLAS Collaboration],
  JHEP {\bf 1503}, 088 (2015)
  [arXiv:1412.6663 [hep-ex]].  
  
  

   
    
  

  
  
  
\end{thebibliography}
\end{document}